\documentclass{article}
\usepackage{makecell} 
\usepackage[table]{xcolor} 
\definecolor{mycolordark}{HTML}{E6E5DC}   
\definecolor{mycolor}{HTML}{FAFAF8}       
\definecolor{mycolorlight}{HTML}{EFEEE9}  

\usepackage{arxiv}
\usepackage{threeparttable}
\usepackage[utf8]{inputenc}
\DeclareUnicodeCharacter{202F}{\,}  

\usepackage[T1]{fontenc}    
\usepackage{hyperref}       
\usepackage{url}            
\usepackage{booktabs}       
\usepackage{amsfonts}       
\usepackage{nicefrac}       
\usepackage{microtype}      
\usepackage{lipsum}
\usepackage{amsmath}
\usepackage{graphicx}
\graphicspath{ {./images/} }
\usepackage[titletoc]{appendix}  
\usepackage{titletoc}

\title{Economic Distance Structures Urban Mobility in 109 U.S. Cities}

\author{
Wei-Peng Nie* \\
  School of Systems Science \\
  Beijing Jiaotong University \\
  Beijing 100044, China \\
  \texttt{wpnie@bjtu.edu.cn} \\
\And
Xiao-Yong Yan \\
  School of Systems Science \\
  Beijing Jiaotong University \\
  Beijing 100044, China \\
  \texttt{yanxy@bjtu.edu.cn} \\
\And
Bin Jia \\
  School of Systems Science \\
  Beijing Jiaotong University \\
  Beijing 100044, China \\
  \texttt{bjia@bjtu.edu.cn} \\
\And
Fang Zhou\\
 Hangzhou Innovation Institute\\
 Beihang University,   \\
 Hangzhou 310052, China \\
  \texttt{zhoufang1995@buaa.edu.cn} \\
\And
Er-Jian Liu \\
    Beijing Transport Institute\\ 
    Beijing 100073, China \\
  \texttt{erjian@bjtu.edu.cn} \\
\And
Tao Zhou* \\
  Comple$\chi$ Lab, Big Data Research Center \\
  University of Electronic Science and Technology of China \\
  Chengdu 610054, China \\
  \texttt{zhutou@ustc.edu} \\
\And
Zi-You Gao* \\
  School of Systems Science \\
  Beijing Jiaotong University \\
  Beijing 100044, China \\
  \texttt{zygao@bjtu.edu.cn}
  }

\begin{document}
\maketitle
\begin{abstract}
Urban mobility promises social integration, yet daily movement is systematically constrained by socioeconomic hierarchies. Introducing "economic distance"—the continuous income gap between origin and destination—as a unified lens, we analyze large-scale mobility records across 109 U.S. cities to reveal how urban flows are structured. We identify a universal structural boundary: flows concentrate intensely within a narrow economic distance of $\sim$0.25 quantiles, defining the effective “economic radius” of routine mobility. This boundary exhibits profound asymmetry; upward mobility faces a uniform structural ceiling across cities, whereas downward mobility drives cross-city heterogeneity. Mechanistically, the boundary is physically anchored by meso-scale residential clustering but is further tightened by an independent economic-distance friction, validated via gravity modeling. These interactions yield four distinct mobility regimes, with “affluent-confined” systems exhibiting the strongest stratification. These findings establish economic distance as a fundamental, asymmetric, and multi-scale filter shaping urban inequality, offering new theoretical grounds for interventions targeting structural barriers to cross-class interaction.

\end{abstract}


\section{Introduction}

Cities aggregate populations, resources, and opportunities, creating spatial proximity that theoretically fosters social mixing and equitable opportunity exchange~\cite{cheng2013measuring, bastiaanssen2022does}. However, geographic proximity does not equate to social interaction. Urban socioeconomic hierarchies continuously reshape human movement, constraining access to urban spaces and daily social engagement~\cite{xu2020deconstructing, yabe2025behaviour}. Uneven routine mobility reproduces systemic social disparities, turning urban agglomeration into a major source of persistent socioeconomic stratification rather than social integration~\cite{torrats2021using, hilman2022socioeconomic}. Understanding the relation between urban socioeconomic configurations and daily mobility patterns is thus essential for revealing the formation of dynamic urban inequality.

For decades, urban stratification research has centered on residential segregation, which describes the static spatial separation of socioeconomic groups across neighborhoods~\cite{massey1990american, williams2001racial}. Classic metrics measuring evenness, clustering and exposure have firmly linked residential sorting to long-term disparities in education~\cite{quillian2014does}, health~\cite{bonal2019residential} and overall socioeconomic well-being~\cite{yilmaz2023access}. Still, this static perspective overlooks the dynamic nature of urban life: residential locations define fixed home locations, while daily mobility governs the full set of urban spaces residents actually experience~\cite{wang2018urban,renninger2025us,athey2021estimating}. Driven by the proliferation of large-scale human trajectory data, a growing body of scholarship has shifted analytical focus from static residential snapshots to dynamic mobility behaviors and their stratified social outcomes~\cite{xu2019quantifying,park2021we}, covering daily travel profiles~\cite{li2023quantifying}, activity space boundaries and spontaneous social encounters~\cite{moro2021mobility, nilforoshan2023human}. 
A robust, cross-city consensus has emerged: individuals disproportionately interact with others from similar socioeconomic backgrounds and travel within socioeconomically familiar neighborhoods~\cite{cook2024urban, zhou2025varying}. While these observations confirm the existence of mobility segregation, they largely inherit the methodological tradition of residential research by discretizing the population into coarse categorical bins (e.g., high/low income). This approach provides only discrete, categorical accounts of socioeconomic sorting, but fails to quantify how the continuous socioeconomic gradient itself structures the flow. Consequently, a fundamental question remains underexplored: does the income gap between origin and destination—defined here as "economic distance"—act as a fundamental organizing force that shapes
the distribution of mobility flows, and at what scale does this force cease to be effective?

Addressing this question requires overcoming several methodological and theoretical limitations in existing literature. First, the prevailing practice of discretizing continuous socioeconomic gradients into categorical bins~\cite{hilman2022socioeconomic,moro2021mobility} artificially erases incremental income differences and precludes the possibility of quantifying the continuous decay of mobility probability with increasing economic distance. Second, current analytical models implicitly assume symmetric mobility frictions along socioeconomic hierarchies, overlooking divergent constraints governing upward (lower-income to higher-income) and downward (higher-income to lower-income) cross-income travel~\cite{xu2019quantifying}. This symmetric assumption masks directional asymmetric mechanisms that mediate cross-group exposure and opportunity access. Furthermore, existing empirical evidence remains fragmented, rarely integrating residential spatial preconditions, micro-level mobility frictions, and cross-city contextual heterogeneity into a unified framework. Together, these limitations hinder a systematic understanding of how economic distance structurally shapes urban mobility.

In this study, we leverage large-scale mobility records across 109 U.S. cities to establish a continuous economic-distance framework for analyzing how urban mobility is structured along economic gradients. We introduce a cumulative flow function to quantify how human flows are distributed along the full socioeconomic gradient, revealing a pervasive short-distance concentration with a universal structural boundary at an economic distance of $\sim$0.25 quantiles. We then unpack the profound directional asymmetry in these constraints: upward mobility faces a structural ceiling across all cities, while downward mobility drives cross-city heterogeneity. We trace the spatial determinants of these patterns, linking them to meso-scale residential clustering, and validate economic distance as an independent micro-level friction via gravity modeling. Finally, we identify four distinct mobility regimes based on these structural features. This work establishes economic distance as a fundamental, asymmetric, and multi-scale filter shaping urban mobility, bridging the long-standing disconnect between static residential sorting and dynamic mobility-driven inequality.

\begin{figure}[t]
    \centering
    \includegraphics[width=1\linewidth,trim=0 0 0 0, clip]{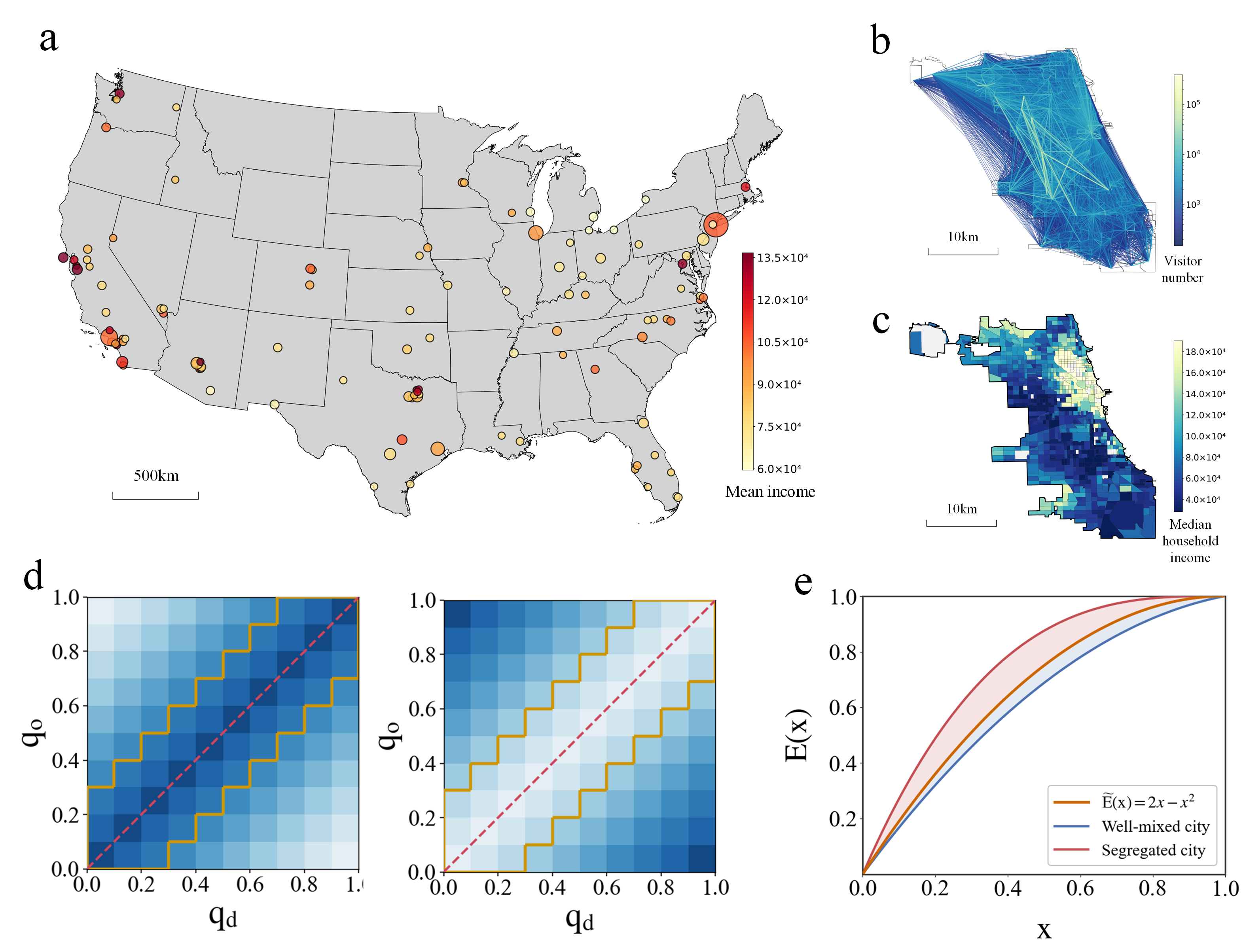}
    \caption{
\textbf{(a)} Map of the contiguous United States (Lower 48), showing the cities included in this study as points. 
Point sizes are proportional to total city population, and point color represents median household income. 
\textbf{(b)} Selected inter-tract mobility flows in Chicago, where line color encodes flow magnitude, highlighting concentrated and sparse flows. 
\textbf{(c)} Census tracts within Chicago colored by median household income, illustrating spatial heterogeneity in socioeconomic status. 
\textbf{(d)} Schematic illustration of two income-ordered OD matrices in income-quantile space: the left panel shows flows concentrated along the main diagonal (high segregation), and the right panel shows flows spread away from the main diagonal (well-mixed). Yellow edges highlight flows within a 0.3 quantile distance from the anti-diagonal.
\textbf{(e)} Corresponding income-distance cumulative flow function $E(x)$: the income-neutral curve $\tilde{E}(x) = 2x - x^2$ (yellow curve) represents income-neutral mixing; the well-mixed city (blue curve) lies below the income-neutral curve, while the highly segregated city (red curve) features rapid accumulation of flows between economically similar neighborhoods. Shaded areas indicate deviations from the income-neutral benchmark.}
    \label{fig:square}
\end{figure}

\section{Results}

We analyze anonymized mobile phone–based mobility data across 109 U.S. cities (see Fig.~\ref{fig:square}a for the geographic distribution of sampled cities), covering daily inter-tract origin–destination flows from January 2019 to February 2020 (see Methods). To capture stable structural mobility patterns, we aggregate these daily flows into a single cumulative OD matrix for each city, representing total inter-tract mobility over the full observation window; Fig.~\ref{fig:square}b illustrates example inter-tract flows in Chicago.
Census tract-level median household income serves as our proxy for socioeconomic status (SES); Figure~\ref{fig:square}c maps its spatial distribution for Chicago. Within each city, we conduct rank-based quantile transformation on census tracts using household income (see Methods). This procedure assigns every tract a continuous income quantile $q_i \in [0,1]$, where $q=0$ represents the lowest-income tracts and $q=1$ represents the highest-income tracts. By doing so, we retain the relative income hierarchy while removing variations in absolute income across different cities. For all subsequent analyses, we operationalize economic distance as the absolute difference in income quantiles between origin and destination tracts, formally defined as income distance: $|q_o - q_d|$. We then sort the OD matrix by income quantiles to obtain income-ordered OD matrices (see Methods).

In these matrices, flows between economically similar tracts ($q_o \approx q_d$) appear near the diagonal, reflecting homophilous interactions. By contrast, flows far from the diagonal connect neighborhoods with large income differences. Accordingly, matrices with flows concentrated along the diagonal (left panel in Fig.~\ref{fig:square}d) represent pronounced flow concentration along economic distance, a structural pattern that corresponds to strong socioeconomic stratification. Matrices with flows spread out from the diagonal (right panel) reflect greater cross-income mixing and weaker stratification. We quantify such structural flow patterns via the income-distance cumulative flow function $E(x)$, defined as the share of flows between tracts whose income distance $|q_o - q_d|$ is smaller than $x$ (the region enclosed by the yellow lines in Fig.~\ref{fig:square}d; see Methods). We compare the observed $E(x)$ curves against an income-neutral benchmark $\tilde{E}(x)=2x-x^2$, which characterizes a fully neutral mixing pattern independent of income (Fig.~\ref{fig:square}e, see Methods). With all metrics established, we proceed to analyze economic-distance structured mobility and its associated socioeconomic stratification across diverse urban contexts.

\begin{figure}[!htbp]
    \centering
\includegraphics[width=0.95\linewidth,trim=5 15 0 12, clip]{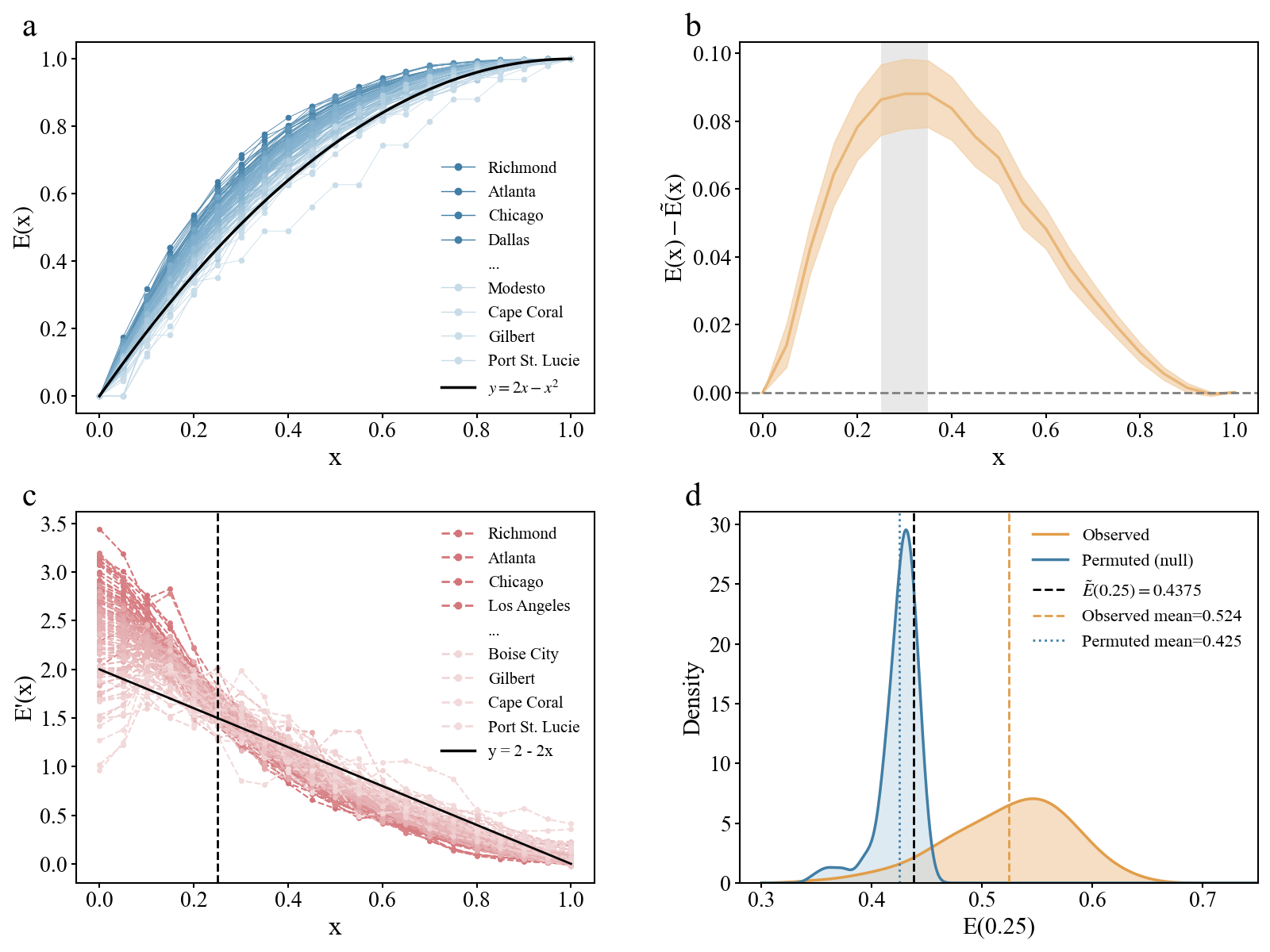}
    \caption{\textbf{(a)} Income-distance cumulative flow functions $E(x)$ for all cities. 
    The majority of curves lie above the income-neutral benchmark (black curve), 
    indicating concentration of mobility within short income distances.
    \textbf{(b)} Mean deviation of cumulative flow functions relative to the income-neutral benchmark, $E(x)-\tilde{E}(x)$, across 109 cities. The solid line shows the average deviation at each income-distance $x$, while the shaded band represents the $95\%$ confidence interval across cities. The average deviation reaches its maximum within the range $[0.25,0.35]$ (highlighted by the gray shaded area). 
    \textbf{(c)} Smoothed derivatives of $E(x)$, with the income-neutral derivative 
    $y=2-2x$ shown in black. Most cities exhibit excess flow density at small income distances ($x<0.25$) relative to the income-neutral benchmark. A vertical dashed blue line at $x=0.25$ marks this characteristic scale. \textbf{(d)} Distribution of the $E(0.25)$ index across 109 cities. The black dashed line marks the income-neutral benchmark $\tilde{E}=0.4375$ and the blue dashed line marks the mean value $0.53$. The observed distribution is shifted far to the right of the permuted distribution (mean 0.425), revealing stronger short-distance mobility concentration than expected under random flow reshuffling.} 
    
    \label{fig:Ecurve}
\end{figure}

\subsection{Universal concentration of flows at short economic distances}

The income-distance cumulative flow function $E(x)$ provides a continuous representation of how urban mobility is organized along the economic-distance gradient, from closely matched to distant neighborhoods. Across all 109 cities, most curves lie systematically above the income-neutral benchmark $\tilde{E}(x)=2x-x^2$ throughout the entire range of $x$ (Fig.~\ref{fig:Ecurve}a). The consistent upward deviation of $E(x)$ relative to the income-neutral benchmark suggests that, across the city, mobility is skewed toward shorter economic distances, with a larger fraction of flows connecting economically closer neighborhoods than would be expected under income-neutral mixing. This systematic upward deviation reveals a universal form of economic-distance structured mobility, in which daily flows are disproportionately concentrated among neighborhoods separated by short economic distances. As a result, mobility systems exhibit a strong tendency toward income-localized interactions, limiting cross-class spatial exposure, which constitutes the macroscopic foundation of urban mobility stratification..

We further quantify the magnitude of this structural flow concentration by calculating the city-level deviation \(E(x)-\tilde{E}(x)\) and the flow density derivative \(E'(x)\). The average deviation rises sharply at short economic distances, peaks at \(x\approx0.25\), and enters a plateau across the 0.25–0.35 range, indicating that excess short-distance flow accumulation reaches its maximum intensity at this narrow economic scale (Fig.~\ref{fig:Ecurve}b). The derivative \(E'(x)\), which resolves fine-grained flow density along the income hierarchy, reveals a sharp structural peak at \(x< 0.25\) (Fig.~\ref{fig:Ecurve}c). Empirical density values substantially surpass the income-neutral benchmark \(y=2-2x\) within this range, confirming extreme over-concentration of flows across minimal economic differences. Crucially, this peak-and-plateau pattern marks a clear structural boundary: economic distance acts as a strong confining force up to $~0.25$ quantiles, but its marginal frictional effect rapidly diminishes beyond this threshold. Consistently, empirical flow densities decline rapidly beyond this threshold and fall below the null benchmark, demonstrating a systemic scarcity of cross-gradient, long-distance socioeconomic mobility. Together, these patterns verify that urban mobility is fundamentally structured by economic distance, revealing a universal "economic social radius" defined by a quantile distance of $~0.25$.

Building on this identified structural boundary, we anchor all subsequent analyses on the representative metric \(E(0.25)\). This threshold captures the core intensity of economic-distance-structured flow concentration, and it delineates the effective economic reach of residents' routine mobility, separating structured homophilous interactions from more random cross-income mixing. Critically, \(E(0.25)\) serves a dual conceptual role that unifies our entire analytical framework. 
Primarily, it acts as a core quantitative indicator characterizing the degree of flow concentration along the economic distance gradient. Secondarily, because such short-range flow concentration inherently restricts cross-class mobility exposure, \(E(0.25)\) reliably reflects emergent mobility stratification across cities. Relative to the income-neutral benchmark \(\tilde{E}=0.4375\) at \(x=0.25\), nearly all cities (101/109) exhibit elevated values, with a right-skewed distribution centered at 0.53 (modal value $\approx$ 0.57; Fig.~\ref{fig:Ecurve}d). Permutation null tests further confirm that observed flow concentration far exceeds randomized flow expectations (Fig.~\ref{fig:Ecurve}d), verifying that economic distance imposes systematic structural constraints on urban mobility. Strong pairwise correlations (\(r>0.95\)) between \(E(0.25)\) and alternative scales (\(E(0.30)\), \(E(0.35)\)) as well as the integrated area under the \(E(x)\) curve further validate its robustness as a representative proxy (Supplementary Section S2). We therefore employ \(E(0.25)\) consistently throughout subsequent analyses to quantify city-level variation in economic-distance structured mobility and its emergent stratification outcomes.

\subsection{Directional asymmetry in economic-distance mobility constraints}

\begin{figure}[!t]
\centering
\includegraphics[width=1\linewidth]{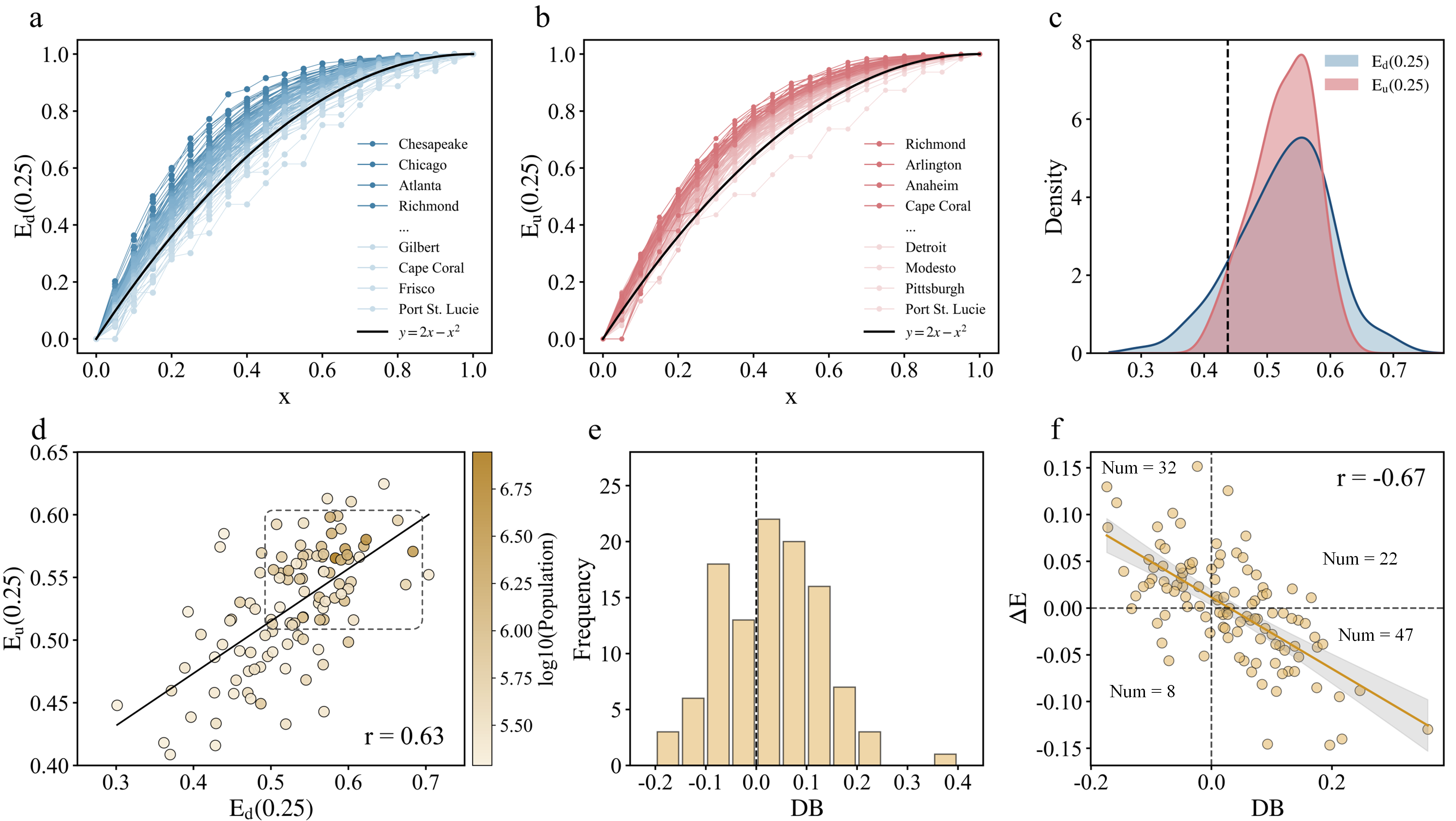}
\caption{ \textbf{(a)} Cumulative curves $E_{\mathrm{d}}(x)$ for downward flows (higher-income to lower-income). 
    The wide dispersion indicates substantial cross-city heterogeneity in segregation in this direction.
    \textbf{(b)} Cumulative curves $E_{\mathrm{u}}(x)$ for upward flows (lower-income to higher-income). 
    Curves are tightly clustered and consistently above the benchmark, reflecting uniformly restricted upward economic mobility.
    \textbf{(c)} Distributions of the directional segregation indices $E_{\mathrm{u}}(0.25)$ and $E_{\mathrm{d}}(0.25)$. 
    Upward indices cluster tightly above the benchmark $\tilde{E}=0.4375$, while downward indices span a broad range.
    \textbf{(d)} Scatter plot of upward segregation index $E_{\mathrm{u}}(0.25)$ versus downward segregation index $E_{\mathrm{d}}(0.25)$ across 109 cities (Pearson $r=0.63$). Point color represents city population. The dotted rectangle highlights the ten largest cities, all of which fall in the high-segregation region of the plot.
    \textbf{(e)} Histogram of directional bias (DB). Positive values denote net downward flow dominance. The distribution is right‑skewed, with $63.3\%$ of cities having DB $>0$.
    \textbf{(f)} Scatter plot of directional bias (DB) against 
    $\Delta E = E_{\mathrm{d}}(0.25)-E_{\mathrm{u}}(0.25)$. 
    Dashed lines mark $x=0$ and $y=0$, dividing the plane into four quadrants.  
    The positive correlation ($r=0.67$) and the asymmetric distribution across quadrants 
    (72.5\% of cities lie in quadrants II and IV) reveal a consistent coupling mechanism: 
    the dominant directional flow “pulls” the opposite direction toward short income distances.}
\label{fig-up-down}
\end{figure}

Having established the universal pattern of short-distance flow concentration across all urban mobility, we next examine whether the structural constraints imposed by economic distance operate uniformly across different travel directions. We decompose overall mobility into two directional components aligned with the income hierarchy: downward mobility (from higher-income to lower-income neighborhoods) and upward mobility (from lower-income to higher-income neighborhoods). These two directional categories correspond to the two off-diagonal flow regimes within the income-ordered OD matrix (see Methods). For each city, we define two directional cumulative flow functions, $E_{\mathrm{d}}(x)$ for downward mobility and $E_{\mathrm{u}}(x)$ for upward mobility. Analogous to the general function $E(x)$, two directional functions quantify the cumulative distribution of directional mobility across the economic distance gradient. 

A direct comparison of $E_{\mathrm{d}}(x)$ and $E_{\mathrm{u}}(x)$ curves reveals pronounced asymmetry in their underlying distributions. Downward cumulative curves exhibit substantially greater heterogeneity across cities, with some falling well below the benchmark and some reaching very high values, whereas upward curves are more tightly clustered and consistently above the benchmark (Fig.~\ref{fig-up-down}a–b). This directional divergence in overall flow pattern is further quantified using the directional mobility structure indicators $E_{\mathrm{d}}(0.25)$ and $E_{\mathrm{u}}(0.25)$ (Fig.~\ref{fig-up-down}c). The two indicators share nearly identical mean values ($\approx0.53$), pointing to comparable average levels of flow concentration across directions. However, their dispersion differs dramatically: $E_{\mathrm{d}}(0.25)$ spans a much wider range ($SD = 0.073$), whereas $E_{\mathrm{u}}(0.25)$ is tightly concentrated around the mean ($SD=0.048$). This pattern demonstrates that economic distance imposes a uniformly strong structural constraint on upward mobility across cities, while the configuration of downward mobility varies substantially across urban contexts. The divergent features in turn produce asymmetric directional mobility stratification as a secondary outcome.

The city-level overall mobility structure, captured by $E(0.25)$, is strongly correlated with both directional mobility structure indicators. Notably, it correlates more strongly with downward mobility ($r=0.94$) than with upward mobility ($r=0.85$) (see Supplementary Fig. S1). This difference indicates that cross-city variation in the overall mobility structure is predominantly linked to downward mobility patterns, whereas upward mobility structure presents far less inter-city variability. Furthermore, $E_{\mathrm{d}}(0.25)$ and $E_{\mathrm{u}}(0.25)$ show a clear positive association ($r=0.63$; Fig.~\ref{fig-up-down}d): cities with stronger short-distance flow concentration in one direction also tend to have elevated flow concentration in the other. Consistent with this shared tendency, large metropolitan areas consistently show intense short‑distance concentration for both upward and downward flows (Fig.~\ref{fig-up-down}d). This suggests that the strong structural constraints imposed by economic distance in large cities affect both downward and upward mobility simultaneously.

To further characterize the directional imbalance in economic-distance structured mobility, we introduce the directional bias index ($\mathrm{DB}$, see Methods) to quantify the net dominance of either downward or upward mobility structure within each city. Positive $\mathrm{DB}$ values indicate that downward mobility dominates the overall flow structure, while negative values signal predominant upward mobility. The city-level distribution of $\mathrm{DB}$ is heavily skewed toward positive values (Fig.~\ref{fig-up-down}e): $63.3\%$ of cities are dominated by downward mobility, with the remainder dominated by upward mobility. We further find a strong statistical association between $\mathrm{DB}$ and the gap in short-distance flow concentration between the two directions, captured by $\Delta E = E_{\mathrm{d}}(0.25)-E_{\mathrm{u}}(0.25)$ (see Supplementary Section S3). The two measures are negatively correlated ($r=-0.67$; Fig.~\ref{fig-up-down}f), and their signs are opposite in $72.5\%$ of sampled cities. When upward mobility becomes the dominant component ($\mathrm{DB} < 0$), downward mobility exhibits stronger short-distance concentration ($\Delta E >0$); conversely, dominant downward mobility ($\mathrm{DB} > 0$) coincides with intensified short-range clustering of upward mobility ($\Delta E <0$). This co-variation reveals a clear gravitational mechanism governing directional mobility structures: the dominant flow component disperses more freely across economic layers—often stretching beyond the typical 0.25 boundary—forming a relatively loose structure, while the minority counter-flow is compressed into narrow short-distance interactions—reinforcing the clustering within the boundary. This gravitational compression holds regardless of which income tier dominates urban mobility, reflecting a universal structural adjustment of daily travel patterns shaped by economic distance.

\subsection{Residential foundations of economic-distance mobility structure}

Prior analyses have established the universal structural boundary at $\sim$0.25 quantiles and its asymmetric constraints on mobility. A critical follow-up question concerns the spatial origins of this structure: does the economic-distance boundary simply mirror the physical separation of income groups, or does it operate through specific spatial scales? 
To address this question, we connect urban mobility structure to two classic theoretical dimensions of residential segregation: evenness–centralization and clustering–exposure~\cite{brown2006spatial}. The evenness–centralization dimension describes macro-level distributional uniformity of income groups across census tracts, while the clustering–exposure dimension characterizes local spatial adjacency and agglomeration of socioeconomically similar neighborhoods.

\begin{figure}[!t]
    \centering
    \includegraphics[width=1\linewidth, trim=0 0 0 0, clip]{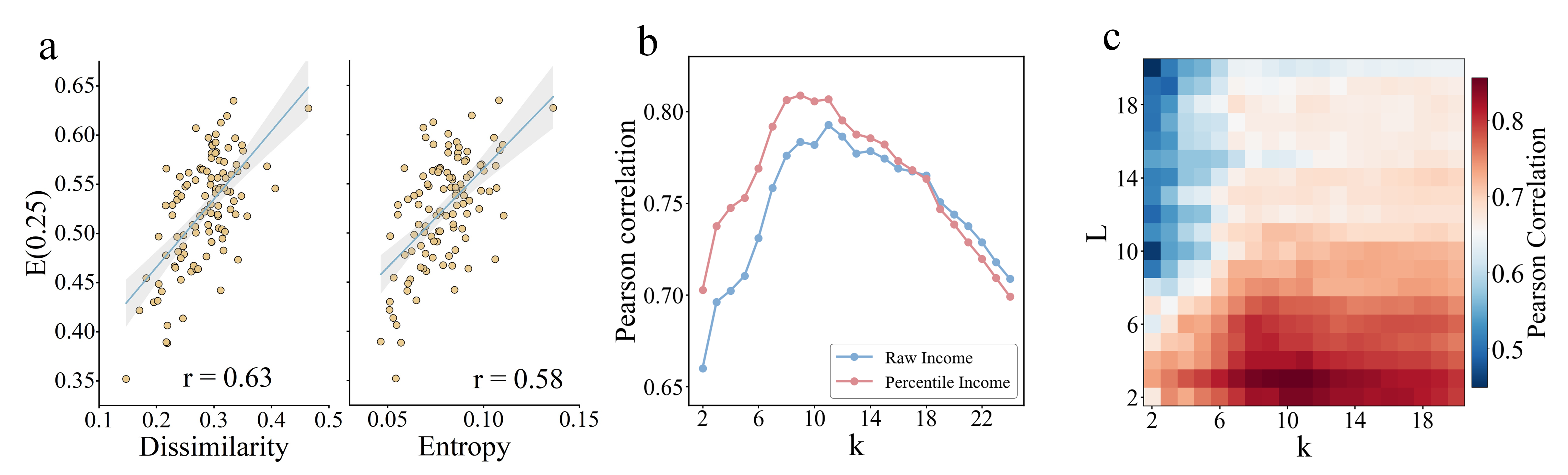}
    \caption{\textbf{(a)} Correlations between $E(0.25)$ and the Dissimilarity Index and Entropy Index. 
    \textbf{(b)} Correlation between $E(0.25)$ and Moran's $I$ across $k$-nearest-neighbor definitions. Using income quantiles yields substantially stronger correlations, which peak at $k=9$ ($r=0.82$). 
    \textbf{(c)} Heat map of correlations between $E(0.25)$ and standardized Join Counts across neighborhood size $k$ and category number $L$. The strongest correlation ($r=0.86$) occurs at $L=3$ and $k=10$.}
    \label{fig:placeholder}
\end{figure}

Correlation analysis reveals clear hierarchical relationships between residential spatial features and urban mobility structure. Indicators for the evenness–centralization dimension, including the Dissimilarity Index and Entropy Index (see Supplementary Section S4), show moderate positive correlations with $E(0.25)$ ($r=0.63$ and $r=0.58$, respectively). This city-wide uneven distribution of income groups provides a weak background constraint for mobility stratification, yet it cannot fully explain cross-city differences in economic-distance flow patterns.

By comparison, the clustering–exposure dimension demonstrates far stronger linkage to mobility structure. We calculate Global Moran’s $I$ using both raw tract median incomes and relative income quantiles. Quantile-based Moran’s $I$, which captures a neighborhood’s rank within the urban income hierarchy, correlates much more strongly with $E(0.25)$ than absolute income values (Fig.~\ref{fig:placeholder}b). This confirms that urban mobility is governed primarily by relative socioeconomic status rather than absolute income levels. The correlation peaks at a $k$-nearest-neighbor threshold of $k=9$ (a flexible approach that extends the neighborhood boundaries of conventional Moran’s I; see Supplementary Section S4), corresponding to a typical meso-scale range of residents’ routine activity spaces. This scale alignment suggests that the “structural boundary” captured by $E(0.25)$ in the economic space finds its physical counterpart in the meso-scale residential clustering: the homogeneity of the $\sim$9 nearest neighbors sets the physical stage for the short-range economic flow concentration.
This pattern suggests that economic-distance structured mobility is most strongly associated with socioeconomic homogeneity at meso-scale spatial extents, rather than with either immediate-neighbor configurations or city-wide income distributions.
We further verify this pattern using standardized Join Counts, which measure excess adjacency of similar-income tracts against random benchmarks (see Supplementary Section S4)~\cite{epperson2003covariances}. The results consistently show the strongest correlation with $E(0.25)$ ($r=0.86$) emerges at $L=3$ income categories and $k=10$ nearest neighbors (Fig.~\ref{fig:placeholder}c), solidifying the conclusion that meso-scale contiguous income clustering is a key residential correlate of economic-distance structured mobility.

Collectively, these results delineate distinct roles for the two residential segregation dimensions. Macro-scale evenness plays a limited secondary role as a city-wide background condition. In contrast, meso-scale clustering acts as the primary spatial generator of the economic-distance boundary, providing the physical template where the structural constraint (reflected by 
$E(0.25)$ takes shape. Overall, these results underscore that the broader residential segregation landscape provides a critical contextual backdrop shaping how economic distance constrains daily urban mobility.

\subsection{Spatial clustering as the dominant driver}

To quantify how much of the observed structural boundary can be attributed to residential sorting versus other factors,  we adopt nested OLS models to quantitatively compare the explanatory power of the two core dimensions of residential sorting. Both dimensions are positively linked to economic-distance mobility stratification: cities with more segregated residential landscapes exhibit stronger stratification in daily mobility patterns (see Column 1 and 2 in Table~\ref{4ols}). However, the clustering–exposure dimension (Moran’s $I$) delivers substantially higher explanatory power than the evenness–centralization dimension ($D$). It accounts for $65\%$ of the variance in $E(0.25)$, compared with $39\%$ for the evenness dimension (adjusted ${R}^{2}=0.649$ vs. $0.387$; $\beta=0.045$ vs. $0.035$, both $p<0.01$). 
This confirms that cross-city differences in economic-distance structured mobility are far better explained by the spatial clustering of income groups than by the overall evenness of residential income distribution. Notably, the high explanatory power ($65\%$) indicates that the structural boundary defined by $E(0.25)$ is predominantly generated by the physical aggregation of income groups at the meso-scale.

Joint regression models reveal that the two residential dimensions capture partially overlapping yet non-redundant variance (Column 3, Table~\ref{4ols}). The coefficient of the evenness metric $D$ declines substantially after controlling for Moran’s $I$, while the coefficient for spatial clustering remains statistically robust. This pattern suggests that a considerable portion of the univariate association between residential evenness and mobility structure overlaps with the variance captured by spatial clustering. Nevertheless, the persistent significance of $D$ indicates that residential evenness provides incremental, unique explanatory power independent of clustering (adjusted ${R}^{2}=0.704$). Further interaction tests identify a statistical substitution relationship between the two dimensions (see Column 4 in Table~\ref{4ols}). The negative and significant interaction term ($\beta=-0.0067$, $p<0.05$) shows that the marginal explanatory contribution of one dimension decreases as the value of the other dimension increases. In statistical terms, cross-city variation in structured mobility tends to be dominated by either clustered or uneven residential spatial features, rather than by a cumulative additive combination of both (adjusted ${R}^{2}=0.719$).

\begin{table}[!t]
\centering
\caption{OLS Regression Results Comparison}
\begin{threeparttable}
\renewcommand{\arraystretch}{1.05}  
\begin{tabular}{lcccc}
\hline
\Xhline{0.3pt} 
\rowcolor{mycolordark}
\rule{0pt}{10pt}
& (1) Univariate ($I$) & (2) Univariate ($D$) & (3) Bivariate & (4) Bivariate with Interaction \\ [2pt] 
\hline
\rowcolor{mycolor} 
\rule{0pt}{10pt} Constant & 0.5238$^{***}$ & 0.5238$^{***}$ & 0.5238$^{***}$ & 0.5273$^{***}$ \\
\rowcolor{mycolor}          & (0.003) & (0.004) & (0.003) & (0.003) \\
\rowcolor{mycolor} Moran's $I$ & 0.0448$^{***}$ & --- & 0.0367$^{***}$ & 0.0347$^{***}$ \\
\rowcolor{mycolor}          & (0.003) &  & (0.003) & (0.003) \\
\rowcolor{mycolor} $D$ & --- & 0.0348$^{***}$ & 0.0156$^{***}$ & 0.0149$^{***}$ \\
\rowcolor{mycolor}          &  & (0.004) & (0.003) & (0.003) \\
\rowcolor{mycolor} $I \times D$ & --- & --- & --- & $-0.0067^{**}$ \\
\rowcolor{mycolor}          &  &  &  & (0.003) \\ [2pt]
\hline
\rowcolor{mycolorlight} \rule{0pt}{10pt} $R^{2}$ & 0.652 & 0.393 & 0.709 & 0.726 \\
\rowcolor{mycolorlight} Adjusted $R^{2}$ & 0.649 & 0.387 & 0.704 & 0.719 \\
\rowcolor{mycolorlight} AIC & $-432.9$ & $-372.2$ & $-450.6$ & $-455.1$ \\
\rowcolor{mycolorlight} BIC & $-427.5$ & $-366.9$ & $-442.5$ & $-444.4$ \\
\rowcolor{mycolorlight} RMSE & 0.0326 & 0.0431 & 0.0298 & 0.0289 \\
\rowcolor{mycolorlight} Observations & 109 & 109 & 109 & 109 \\ [2pt]
\Xhline{0.3pt} 
\hline
\end{tabular}
\begin{tablenotes}
\footnotesize
\item Note: All explanatory variables are standardized. Standard errors in parentheses. * $p<0.1$, ** $p<0.05$, *** $p<0.01$.
\end{tablenotes}
\end{threeparttable}
\label{4ols}
\end{table}

In total, residential spatial dimensions account for approximately $73\%$ of cross-city variation in economic-distance mobility structure. The remaining $\sim 27\%$ of unexplained variance carries substantive structural information, suggesting that economic-distance constraints are not merely a passive mirror of residential patterns but involve additional mechanisms. The residual variance exhibits weak but systematic statistical associations with two urban socio-demographic attributes (see Supplementary Table~S3)~\cite{lovell2008simple}. Racial diversity is positively correlated with excess short-distance flow concentration in mobility ($\beta = 0.0086$, $p=0.006$), indicating that racially diverse cities display subtle mobility structural deviations beyond the constraints imposed by residential spatial patterns. In contrast, a larger share of residents under age 18 correlates negatively with such excess concentration ($\beta = -0.0114$, $p=0.003$), corresponding to more mixed, less structurally constrained daily mobility patterns in younger cities.

Consistent with the directional asymmetry observed in prior analyses, the explanatory power of residential spatial structure varies notably across mobility directions (see Supplementary Table~S4). Residential dimensions explain a considerably larger proportion of variance in downward mobility structure (adjusted ${R}^{2}=0.658$) compared with upward mobility structure (adjusted ${R}^{2}=0.498$), suggesting that downward directional mobility patterns are statistically more closely aligned with urban residential configurations. Spatial clustering (Moran’s $I$) also presents stronger predictive coefficients for downward mobility ($\beta=0.0439$, $p<0.001$) than for upward mobility ($\beta=0.0254$, $p<0.001$). This contrast indicates that the downward flows largely follow the path laid by residential clustering, whereas upward mobility is less “structurally determined” by residential patterns alone. This statistical distinction provides a spatial root for the directional asymmetry identified earlier: upward constraints appear to be more sensitive to non-residential frictions.

Residual variance after accounting for residential spatial structure also displays directional heterogeneity. For upward mobility, residual values show marginally negative associations with public transit accessibility ($\beta=-0.0092$, $p=0.099$) and weakly positive associations with average commute duration ($\beta=0.0094$, $p=0.072$). These marginal trends suggest modestly reduced excess flow concentration in upward mobility within cities with better transit access and shorter commutes. For downward mobility, residual variation correlates weakly with racial diversity ($\beta=0.0103$, $p=0.052$) and youth population share ($\beta=-0.0147$, $p=0.019$). Although these residual associations are small in magnitude, their consistent directional patterns reveal a systematic statistical distinction: upward mobility variation tends to co-vary with transportation infrastructure attributes, while downward mobility variation is more closely correlated with urban socio-demographic composition.

\subsection{Heterogeneous Effects across Urban Contexts}

\begin{figure}[!t]
\centering
\includegraphics[width=1\linewidth]{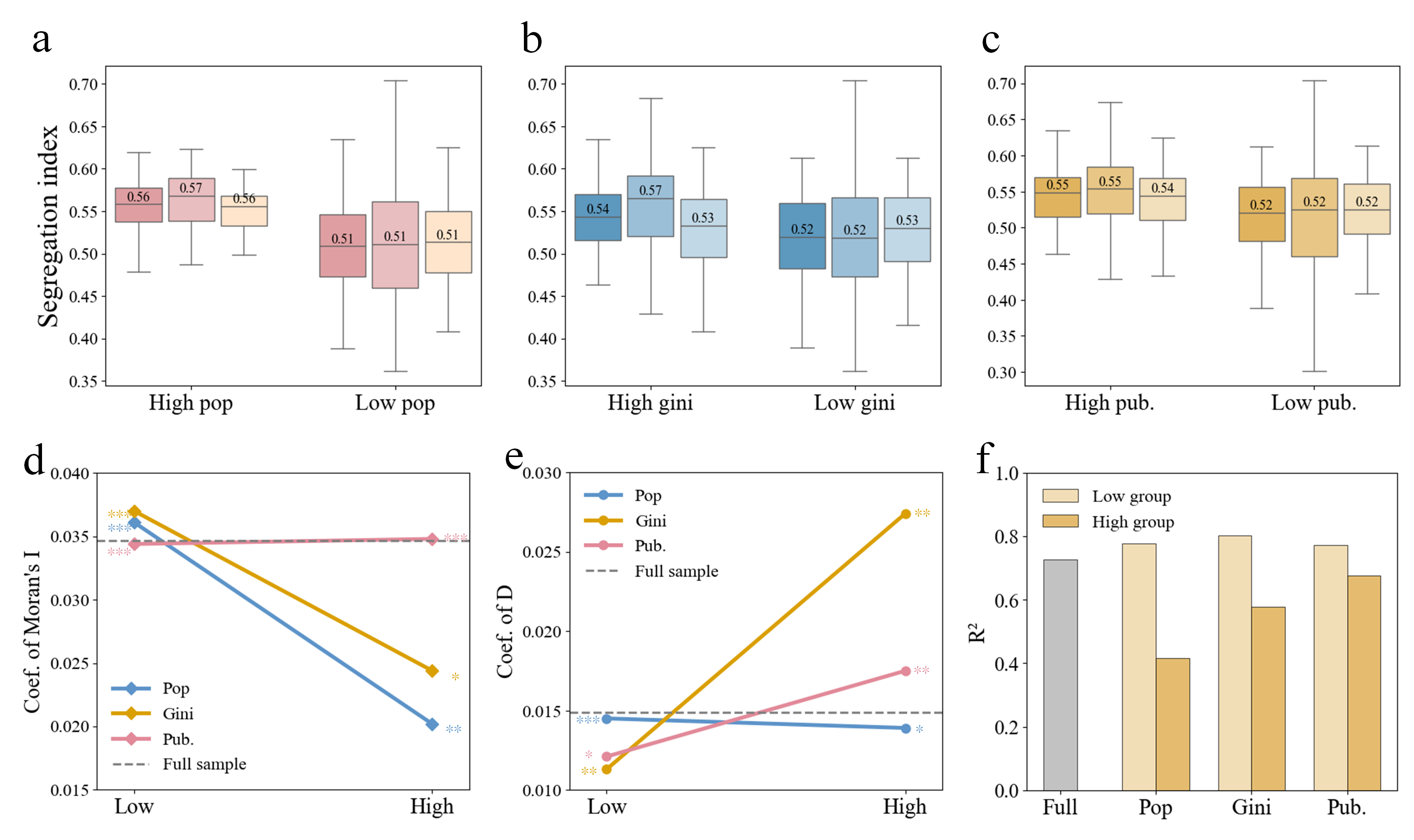}
\caption{\textbf{(a-c)} Mobility segregation indices by urban subgroup. Panels: population size (pop), income Gini (gini), public transit usage (pub.). Each panel splits cities into high (top 1/3) and low (bottom 2/3) subgroups; the three box colors per subgroup represent $E(0.25)$, $E_{\text{down}}(0.25)$, $E_{\text{up}}(0.25)$ from dark to light, respectively. Median values are labeled above boxes. \textbf{(d)} Group‑specific standardized coefficients for Moran’s $I$ in regressions of $E(0.25)$. Points represent low (small/medium cities, low inequality, low transit) and high (large cities, high inequality, high transit) subgroups for each urban dimension, connected by lines. Horizontal dashed line indicates the full‑sample coefficient. Significance markers ($^{}p<0.05$, $^{}p<0.01$, $^{}p<0.001$) are placed above each point.
\textbf{(e)} Same as \textbf{(d)} but for the coefficient of $D$. \textbf{(f)} Model $R^2$ for group-specific OLS regressions of $E(0.25)$.}
\label{fig:hetero_boxplot}
\end{figure}

The preceding analyses establish robust average associations between residential spatial configurations and city-level economic-distance mobility structure. We next extend this framework to examine contextual heterogeneity, evaluating how the statistical linkage between residential sorting dimensions (spatial clustering and income evenness) and structured mobility varies across urban conditions. We stratify cities along three core contextual dimensions: population size, income inequality (Gini index), and public transit usage. For each dimension, cities are grouped into high (top one-third) and low (bottom two-thirds) subgroups based on city-level attribute rankings~\cite{johnson2020lead}. Descriptive boxplots reveal clear disparities concentrated in city size: large cities show consistently higher median mobility segregation with $E(0.25)=0.558$, $E_{\text{down}}(0.25)=0.568$, and $E_{\text{up}}(0.25)=0.556$, while small and medium cities exhibit systematically lower values (Fig.~\ref{fig:hetero_boxplot}a). This suggests that the structural boundary defined by economic distance is more restrictive in larger urban systems. Differences between high/low income inequality and high/low public transit usage are moderate but directionally consistent, with higher groups marginally more segregated overall (Fig.~\ref{fig:hetero_boxplot}b–c).

The regression results reveal a consistent shift in model performance and coefficient patterns across the three contextual dimensions: as population size, income inequality, and public transit usage increase, the explanatory power of residential structure declines substantially, the coefficient of spatial clustering (Moran's $I$) decreases monotonically, and the coefficient of income evenness ($D$) progressively rises (Fig.~\ref{fig:hetero_boxplot}d–f). Minor deviations from this pattern are marginal and do not alter the dominant direction.
Across all subgroups, the negative interaction term between the two residential dimensions remains significant, indicating a persistent substitutive relationship. Notably, the magnitude of change varies across dimensions: the drop in $R^2$ is steepest for city size (from 0.78 to 0.42), while the rise in the $D$ coefficient is most pronounced for income inequality (from 0.015 to 0.027). This distinction shows that each urban dimension modulates the statistical coupling between residential configuration and mobility structure with slightly different intensity. Thus, larger, more unequal, and high-transit-access urban environments weaken the statistical linkage between residential patterns and daily mobility, with the relative predictive weight gradually shifting from spatial clustering toward city-wide income evenness.  This decoupling implies that in complex urban systems, the economic-distance structure 
($E(0.25)$) is no longer solely anchored by immediate physical neighbors but is increasingly shaped by broader socioeconomic distributions across the city.

Beyond the three core urban dimensions examined above, we further test three additional city-level attributes, including population density, racial diversity, and aging pressure (see Supplementary Section S7 for full regression results). Collectively, the six urban indicators fall into two distinct moderating categories based on their impacts on residents’ daily mobility scope. The first category (city size, income inequality, and public transit usage) expands individual activity ranges and decouples mobility from residential constraints, weakening the predictive capacity of residential structure. In contrast, the second category (density, racial diversity, and aging pressure) compresses activity spaces and strengthens local environmental dependence: high density compresses travel horizons, racial diversity may anchor trips within ethnic enclaves, and aging populations have intrinsically shorter activity ranges. For all three attributes in this category, high-group cities exhibit elevated model explanatory power, indicating tighter coupling between residential spatial configurations and mobility structure. 
This bidirectional moderation reveals a nuanced contextual mechanism shaping urban mobility stratification: urban features that broaden mobility relax the structural boundary, allowing flows to transcend local residential constraints, while factors reinforcing local attachment tighten the boundary, anchoring daily mobility to neighborhood conditions.

\subsection{Economic distance as a systematic friction in gravity flows}

The preceding sections have documented how urban residential configurations and contextual attributes correlate with cross-city variation in economic-distance mobility structure. This raises a fundamental question: after accounting for these macro and meso-scale structural factors, does economic distance itself function as an independent force shaping individual travel choices at the micro level? We address this mechanism by adopting two nested gravity model specifications for each city: a baseline gravity model (BGM) and an extended economic-distance gravity model (EGM) that explicitly incorporates the economic-distance term (see Methods).

\begin{figure}[t]
    \centering
    \includegraphics[width=1.0\linewidth,trim=0 25 0 0, clip]{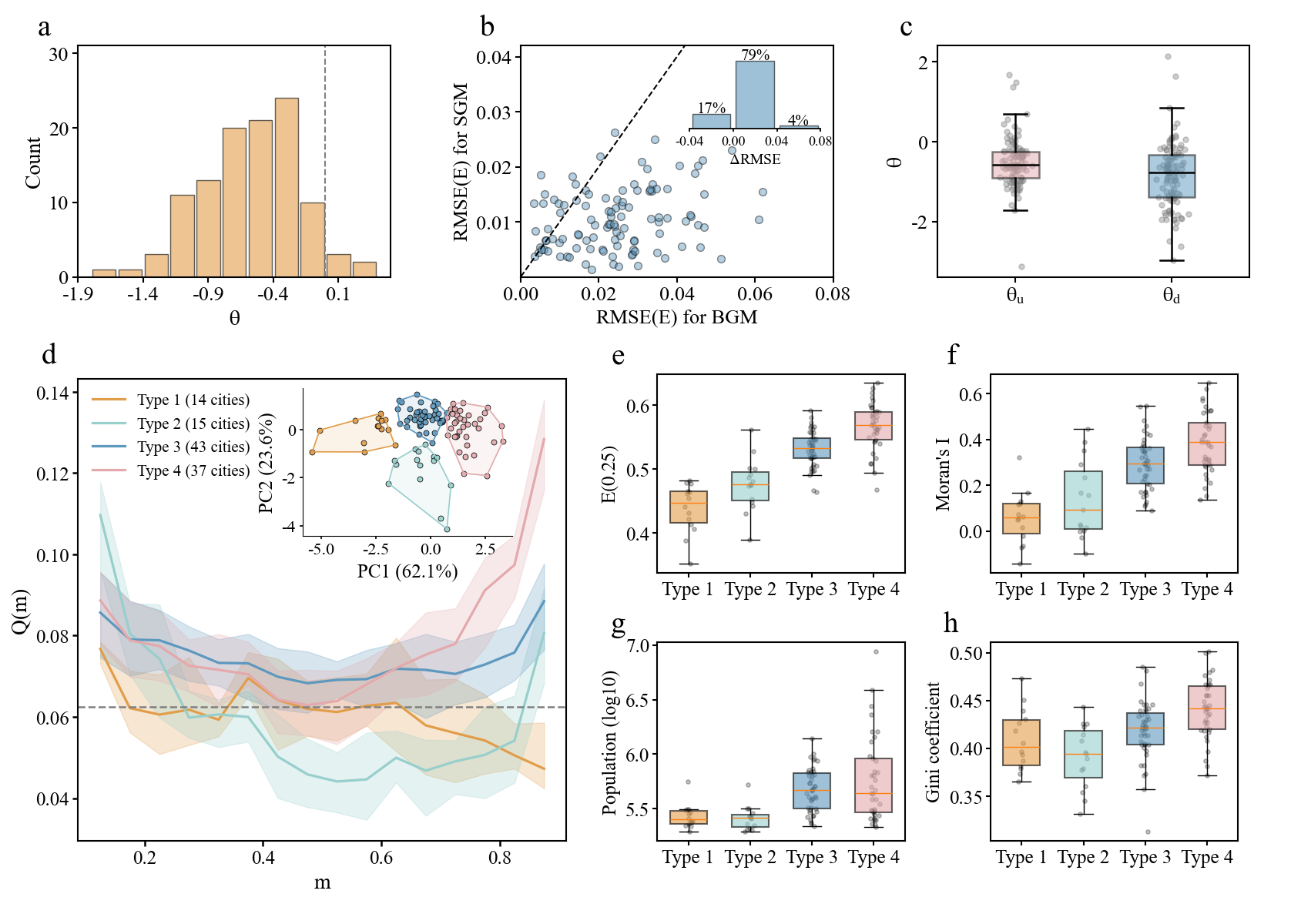}
    \caption{\textbf{(a)} Distribution of the economic-distance coefficient $\theta$ across cities. Majority cities show negative $\theta$ values, which imply reduced mobility flows across larger income gaps, conditional on origin size, destination size, and geographic distance.
\textbf{(b)} Scatter plots of the root mean squared error (RMSE) between observed and predicted Cumulative curves for the $E(x)$ function. Most scatters lie below the diagonal (dashed line, $y=x$). Insets show the distribution of $\Delta$RMSE (BGM-EGM), with negative values indicating improved performance of the EGM. RMSE reductions are observed in the majority of cities ($83\%$), indicating that economic distance systematically improves the reconstruction of mobility structure.
\textbf{(c)} Distribution of economic-distance coefficients ($\theta$) for upward and downward flows. Half-violin plots with overlaid points show that coefficients are more negative for downward flows. \textbf{(d)} Mean $Q(m)$ curves across the four city types, with shaded regions representing interquartile ranges (IQR).
The inset presents the two-dimensional PCA projection of the four clustering features (endpoint intensity, asymmetry index, U-shape depth, and overall mean $\bar{Q}$). Convex hulls delineate the boundaries of four city typologies, demonstrating clear separation in the reduced feature space. PC1 and PC2 explain $62.15\%$ and $23.65\%$ of the total variance, respectively (cumulative: $85.79\%$).
\textbf{(e--h)} Distributions of key city characteristics across the four mobility typologies:
(\textbf{e}) overall mobility segregation $E(0.25)$,
(\textbf{f}) residential spatial clustering (Moran’s $I$),
(\textbf{g}) population size ($\log_{10}$ scale),
and (\textbf{h}) income inequality (Gini coefficient).}
    \label{fig:allgravity-q}
\end{figure}

We first compare the overall performance of the baseline gravity model (BGM) and the economic-distance gravity model (EGM). By model design, the EGM includes an additional economic-distance parameter and thus yields slightly higher $R^2$ values than the BGM across all cities, yet the overall improvement is modest, with a median $R^2$ increment of only 0.01. The estimated economic-distance coefficient $\theta$ quantifies the frictional effect of income gaps on mobility (Fig.~\ref{fig:allgravity-q}a). Across the 109 sampled cities, $\theta$ is negative in $94\%$ of cases (mean $\bar{\theta}=-0.57$, $95\%$ CI: $[-0.64, -0.50]$). This pattern indicates that mobility flows gradually decline as income gaps expand in most cities, after controlling for origin-destination scale and geographic distance. This confirms economic distance as an independent friction force. The negative $\bar{\theta}=-0.57$ quantifies the “economic stickiness” that pulls flows back to their origin income level. At the aggregate city level, this pairwise friction operates alongside—and further reinforces—the spatial constraints imposed by physical distance and residential clustering, contributing to the macroscopic structural boundary ($E(0.25)$) observed earlier. The result is statistically robust: only 9 cities produce non-significant $\theta$ estimates at the $5\%$ significance level.

These findings confirm that economic distance acts as a pervasive, city-wide systematic friction for urban interaction. It universally restricts flows between socioeconomically distinct neighborhoods, rather than exerting sporadic or context-dependent effects. Beyond aggregate model fit, adding the economic-distance term also optimizes the reconstruction of mobility patterns along the income gradient. For $83\%$ of cities (90 out of 109), the EGM produces lower RMSE for the fitted $E(x)$ curve compared with the BGM (Fig.~\ref{fig:allgravity-q}b). The consistent improvements in both $R^2$ and structural RMSE demonstrate that economic distance not only boosts aggregate model performance, but also better captures the internal organizational features of urban mobility structure. 
This provides direct evidence that the micro-level friction captured by $\theta$ operates as one of the generative mechanisms underlying the macro-level flow concentration.

We further explore directional heterogeneity by estimating separate gravity models for upward and downward mobility (see Supplementary Section S8). The economic-distance coefficients reveal a pronounced directional divergence. Coefficients for downward mobility are consistently more negative than those for upward mobility (mean $\theta$: $-0.85$ vs. $-0.52$), meaning flows from higher-income to lower-income neighborhoods experience stronger frictional constraints imposed by income gaps (see Fig.~\ref{fig:allgravity-q}c). A paired t-test verifies this divergence is statistically significant ($t=5.43$, $p<1\times10^{-6}$). This directional pattern aligns well with the asymmetric features of directional mobility structure identified in earlier correlation and regression analyses. Despite such heterogeneity in frictional intensity across directions, the economic-distance term stably improves model performance for both flow types. When splitting mobility by direction, the EGM still reduces the reconstruction error of Cumulative curves relative to the BGM in most cities (see Supplementary Fig.~S2). This outcome proves that economic distance is a generalizable core factor for characterizing structured mobility across all travel directions.

\subsection{Urban mobility regimes shaped by income-localized flow concentration}

The preceding sections established economic distance as a multi-scale filter that generates a universal structural boundary. We now dissect the internal structure of this boundary to understand who is most constrained. While $E(x)$ quantifies the overall intensity of the filter, it masks heterogeneity in how the filtering effect is distributed across the income hierarchy. To resolve this, we introduce the sliding-window measure $Q(m)$, which quantifies the fraction of flows occurring within a narrow income band centered at quantile $m$ (see Methods)

Across cities, the average $Q(m)$ profile exhibits a pronounced U-shaped pattern (Fig.~\ref{fig:allgravity-q}d and Supplementary Fig.~S3), indicating that economically localized mobility is generally concentrated near both ends of the income hierarchy while remaining comparatively weaker among middle-income neighborhoods. Despite this common tendency, substantial heterogeneity exists in the depth, symmetry, and endpoint dominance of these profiles. To characterize such variation, we performed K-means clustering using four summary features derived from each city's $Q(m)$ curve: endpoint intensity, asymmetry, U-shape depth, and overall concentration level (see Methods).

The clustering reveals four distinct regimes representing varying degrees and modes of the economic-distance filter (Fig.~\ref{fig:allgravity-q}d). Type~1 (Flat mixed, 14 cities) represents a weak/porous filter, where flows exhibit weak concentration at both income extremes and a relatively flat profile. Type~2 (Low-concentrated deep-U, 15 cities) and Type~4 (High-concentrated deep-U, 37 cities) represent strong filters but with opposing targets: Type~2 is defined by lower-class enclosure, where low-income neighborhoods exhibit strong localization, while Type~4 is defined by upper-class enclosure, where affluent neighborhoods exert the strongest structural constraint. Type~3 (Symmetric shallow-U, 43 cities) the most common pattern, shows balanced concentration at both ends of the income hierarchy with a moderate U-shape. These four regimes differ substantially across all clustering dimensions (Supplementary Table~S6). Their separation is further supported by PCA projections of the feature space, where cities occupy distinct and non-overlapping regions (Fig.~\ref{fig:allgravity-q}d inset; Supplementary Section S9), suggesting that the identified patterns reflect qualitatively different mobility organizations rather than continuous random variation.

Importantly, these mobility regimes are systematically associated with broader urban characteristics (Fig.~\ref{fig:allgravity-q}e--h). Moving from Type~1 to Type~4, cities exhibit generally higher overall flow concentration captured by $E(0.25)$, stronger residential spatial clustering (Moran's $I$), larger population size, and greater income inequality. Type~4 cities simultaneously display the highest levels of flow concentration and residential clustering, whereas Type~1 cities tend to be smaller, less unequal, and less spatially clustered. Notably, Type~2 and Type~4 cities both exhibit pronounced U-shaped profiles but differ markedly in which end of the income hierarchy dominates localized mobility. While Type~2 cities are characterized by stronger concentration among low-income neighborhoods, Type~4 cities are distinguished by strong affluent-localized mobility together with substantially higher overall flow concentration. Consistent with the directional asymmetry identified earlier, cross-city variation in overall mobility structure aligns more strongly with concentration at the high-income end of the income hierarchy than at the low-income end. This pattern suggests that affluent-localized mobility constitutes a particularly influential component of economic-distance structured mobility and may play a disproportionate role in the emergence of city-level mobility stratification. Additional differences in median income, poverty rate, and educational attainment further indicate that socioeconomic composition contributes to the emergence of distinct mobility regimes (Supplementary Fig.~S4).

Taken together, these results show that the economic-distance filter operates in systematically different modes across cities. Beyond the common tendency toward short-distance concentration, cities exhibit distinct internal organizations of economically localized mobility, ranging from relatively integrated regimes to forms with strong affluent-centered flow concentration and stratification. Notably, the most intense stratification (Type~4) is associated not merely with strong constraints on the poor, but specifically with the “secession” of the affluent into highly localized mobility circuits. This finding refines our understanding of the directional asymmetry: while upward constraints act as a universal ceiling, it is the variability in downward/localized affluent mobility that defines the distinct structural regimes of urban systems. The $Q(m)$ typology therefore extends the aggregate perspective provided by $E(x)$, revealing how a shared structural mechanism—economic-distance mobility concentration—can generate heterogeneous mobility regimes across urban systems.

\section{Discussion and Conclusion}

Urban mobility is widely recognized as a pathway to social integration, yet daily movement is systematically structured along socioeconomic hierarchies~\cite{xu2025using}. Our study introduces economic distance—the income-differential between origin and destination tracts—as a continuous, cross-city comparable lens to quantify how urban flows are organized. We identify economic distance as a fundamental, asymmetric structural filter that constrains urban interaction. A central finding is the identification of a universal “structural boundary” at an economic distance of ~0.25 quantiles. This boundary delineates the effective economic reach of routine mobility, beyond which flow density drops sharply, suggesting that the “economic social radius” is a fundamental property of urban systems.

The operation of this filter is shaped by a hierarchy of mechanisms. At the spatial roots, the boundary is primarily generated by meso-scale residential income clustering (defined by ~9 nearest neighbors). This specific scale alignment—where physical neighbors define the economic boundary—explains why previous studies focusing solely on static residential segregation or immediate neighbors failed to capture the full dynamics. The filter acts as a bridge between static spatial patterns and dynamic behavioral outcomes: the meso-scale homogeneity of neighborhoods sets the physical stage for the 0.25 economic distance concentration.
At the micro-level, gravity modeling confirms that economic distance functions as an independent friction force beyond geographic proximity. This friction acts as a behavioral “tightener”: it prevents flows from diffusing freely even when physical travel costs are low, thereby reinforcing the stratification pattern initiated by residential clustering. Thus, the macroscopic boundary (E(0.25)) emerges as a composite outcome: it is physically framed by residential structure and behaviorally tightened by independent economic friction. 
Furthermore, the filter exhibits profound directional asymmetry. Upward mobility faces a uniform “structural ceiling” across all cities, likely reflecting exclusionary constraints on accessing high-opportunity areas. In contrast, downward mobility drives cross-city heterogeneity, varying from free dispersion to strict confinement depending on urban contexts. This asymmetric pattern cannot be explained by symmetric geographic distance or static residential clustering alone~\cite{Chetty2014,xu2019quantifying}. 

The strength and shape of this economic-distance filter are systematically moderated by urban context and exhibit distinct typologies. Activity-expanding attributes (e.g., city size, transit usage) decouple mobility from local residential constraints, effectively weakening the linkage between physical clustering and the structural boundary. Conversely, features like high density or aging populations tighten the filter, anchoring mobility to neighborhood conditions~\cite{chen2016effects}. The mobility typology analysis further reveals that the most intense stratification (Type 4) is defined by “affluent enclosure,” where the filter is strongest at the high-income end. This suggests that while the filter universally constrains the poor, the structural secession of the affluent is the hallmark of highly stratified systems.

The multiscale heterogeneity observed has distinct policy implications across spatial scales and mobility systems. First, because the meso‑scale residential clustering (~9 nearest neighbours) is the spatial unit generating the boundary, policies should aim to break up income‑homogeneous zones at this right scale—for instance, through mixed‑income housing developments and the creation of shared public spaces that attract diverse groups~\cite{kontokosta2014mixed,lens2016strict}. Second, for large, highly stratified cities, mitigating city‑wide income sorting (e.g., via inclusive zoning or job‑housing balance) becomes more urgent than neighbourhood‑scale interventions. Third, improving public transit networks can partially relieve the uniform upward mobility constraint, especially when combined with affordable housing near high‑opportunity areas~\cite{akbar2025public}. Methodologically, our continuous attribute‑based framework ($E(x)$ index, directional decomposition, sliding‑window $Q(m)$ typology) is generalizable beyond urban mobility. It can be applied to any network where nodes carry continuous attributes (e.g., income, age, education) to reveal hierarchical connectivity and attribute homophily, offering a new tool for inequality research in complex systems.

Nevertheless, several limitations must be acknowledged, along with directions for future research. Our data capture spatial co‑presence but not trip purposes or the quality of social interactions~\cite{luo2026human}. The cross‑sectional design cannot establish long‑term evolutionary dynamics of mobility stratification, nor can it fully rule out unobserved confounding~\cite{hilman2023mobility}. All results are derived from U.S. cities, and cross‑national validation is needed to assess generalizability. Future research should integrate surveys or ethnographic methods to understand the subjective experience of economic‑distance constraints, and should leverage quasi‑experimental designs (e.g., transit line openings, mixed‑income housing lotteries) to strengthen causal inference.

This work identifies economic distance as a multi-scale, directionally asymmetric core rule governing urban mobility structure. By departing from discrete income grouping and conventional symmetric indicators, we show that urban flows are shaped by both geographic proximity and inherent socioeconomic constraints. As a layered structural force, economic distance concentrates flows within narrow income ranges, imposes universal restrictions on upward mobility while generating notable cross-city variation in downward mobility, and retains an independent frictional effect on cross-income travel after controlling for geographic distance and residential configurations. These findings revise our interpretation of urban mobility disparities: such inequalities stem jointly from residential spatial patterns and the sorting driven by economic distance. 

\section{Methods}

\subsection{Data Sources and Preprocessing}

This study integrates three types of data: anonymized mobile phone travel flows, census tract (CT)-level income data, and city-level socioeconomic indicators, forming the empirical basis for analyzing the relationship between urban mobility structure and socioeconomic structure. We use mobile phone signaling data from 109 largest U.S. cities by population, provided in the form of daily origin–destination (OD) matrices at the census tract level~\cite{kang2020multiscale}. Each CT is treated as a basic spatial unit, and OD entries record daily flows from individuals’ home tracts to visited destination tracts. Home locations are inferred as the most frequently visited nighttime tract (6:00\,pm–7:00\,am local time) over a six-week observation window. Daily flows are aggregated at the individual level, counting a person only once per day for each unique destination tract, regardless of the number of visits and intermediate stops. 

Socioeconomic status at the tract level is proxied by median household income, obtained from the U.S. Census Bureau for all CTs within the sampled cities. To characterize broader urban socioeconomic and demographic contexts, we compile a set of city-level indicators covering population composition, education and employment, income and inequality, transportation behavior, and social security. Specifically, these include population size, minority proportion, racial diversity, age structure, educational attainment, occupational structure, unemployment rate, income and poverty measures (median income, per capita income, Gini coefficient, poverty rate), commuting mode shares, average commute time, and health insurance coverage (see Supplementary Section S1).

\subsection{Income Quantile Transformation}
We use tract-level median household income as a proxy for socioeconomic status. To ensure comparability across cities and to avoid differences in absolute income scales, we normalize incomes by rank-ordering tracts within each city. Specifically, for a tract with income $y_i$, its within-city income quantile is defined as:
\begin{equation}
    q_i = \frac{\mathrm{rank}(y_i)-1}{N-1},
\end{equation}
where $\mathrm{rank}(y_i)$ is the average rank of $y_i$ among all tracts in the city (using average ranks for ties), and $N$ is the total number of tracts. This maps incomes monotonically onto the interval $[0,1]$, with $q=0$ and $q=1$ denoting the lowest and highest income tracts, respectively. Importantly, this transformation preserves only the relative income ordering, not absolute income differences.

\subsection{Income-ordered OD matrix and cumulative flow function}

We construct an income-ordered OD matrix by aligning rows (origins) and columns (destinations) by their income quantile $q$. Diagonal entries, which represent intra-tract flows, are removed so that the analysis focuses exclusively on inter-tract mobility. For each OD pair $(o,d)$, we define its income distance as $|q_o-q_d|$, which serves as a tractable proxy for economic distance in urban mobility systems. Let $F_{od}$ denote the (non-diagonal) flow volume from tract $o$ to $d$, 
where flows are normalized by the total outgoing volume of each origin tract, 
so that $\sum_d F_{od} = 1$ for all $o$.  We define the cumulative flow function based on income distance as:
\begin{equation}
    E(x) = P(|q_o-q_d|<x) 
    = \frac{\sum_{o,d} F_{od}\,\mathbf{1}(|q_o-q_d|<x)}{\sum_{o,d} F_{od}},
\end{equation}
where $\mathbf{1}(\cdot)$ is the indicator function (see Fig.~\ref{fig:square}d). Thus, $E(x)$ corresponds to the weighted cumulative distribution function (CDF) of income distances among all flows in the city. This approach follows the general idea of assessing mobility structure through ordered cumulative flows, extending hierarchical flow analyses from activity-based levels to a continuous income-ranked space~\cite{bassolas2019hierarchical}.

\subsection{Income-neutral benchmark}

To establish a common benchmark across cities, we adopt an income-neutral cumulative function that assumes mobility is independent of income distance. Given our flow normalization ($\sum_d F_{od}=1$ for all $o$), this null hypothesis is equivalent to assuming that every origin-destination pair is equally likely. Under this random matching in the income-quantile space, the probability that a random flow has an income distance less than $x$ admits a closed-form expression based on geometric probability:
\[
\tilde{E}(x) = 2x - x^2,\qquad x\in[0,1].
\]
This curve serves as a universal theoretical expectation against which the observed $E(x)$ is compared. Observed curves lying above $\tilde{E}(x)$ indicate an overconcentration of flows at short income distances; such flow concentration further manifests as mobility segregation. Curves falling below the benchmark reflect a higher degree of cross-income flow mixing.

\subsection{Directional decomposition and  directional cumulative flow functions}
\label{subsec:directional}
To test for potential asymmetry in how economic distance constrains mobility, we decompose aggregate flows into movements along the income hierarchy: upward ($q_o < q_d$) and downward ($q_o > q_d$). In the income-ordered OD matrix, upward flows correspond to cells above the diagonal, and downward flows correspond to cells below the diagonal. Define the directional flow components as

\[
F_{od}^u = F_{od}\,\mathbf{1}(q_o < q_d), \qquad 
F_{od}^d = F_{od}\,\mathbf{1}(q_o > q_d).
\]
The total directional flow volumes used for normalization are calculated as:

\[
U = \sum_{o,d} F_{od}^u, \qquad D = \sum_{o,d} F_{od}^d.
\]
The directional cumulative flow functions are then defined as

\begin{align}
    E_u(x) &= \frac{\sum_{o,d} F_{od}^u\,\mathbf{1}(|q_o-q_d|<x)}{U}, \label{eq:Eu} \\
    E_d(x) &= \frac{\sum_{o,d} F_{od}^d\,\mathbf{1}(|q_o-q_d|<x)}{D}. \label{eq:Ed}
\end{align}
$E_u(x)$ and $E_d(x)$ represent, respectively, the proportion of upward and downward flows that occur within an income distance smaller than $x$. They share the same numerator structure as $E(x)$ but are normalized by the corresponding directional totals, allowing us to compare concentration patterns independently for each direction. As with the aggregate function $E(x)$, the income-neutral benchmark for these directional cumulative flow functions is $\tilde{E}(x)=2x-x^2$, corresponding to income-neutral mixing in the income-quantile space. 

\subsection{Directional bias index}
To quantify the net directional orientation of mobility flows along the income hierarchy, we define the directional bias index as:
\[
\mathrm{DB} = \frac{D_{\text{raw}} - U_{\text{raw}}}{U_{\text{raw}} + D_{\text{raw}}},
\]
where \(U_{\text{raw}}\) denotes the total volume of upward flows (from lower‑income to higher‑income tracts) and \(D_{\text{raw}}\) denotes the total volume of downward flows (from higher‑income to lower‑income tracts). Both \(U_{\text{raw}}\) and \(D_{\text{raw}}\) are computed using the original (non‑normalized) flow weights, and only flows connecting tracts with distinct income levels are considered (i.e., flows between tracts belonging to the same income group are excluded).

\subsection{Gravity Model Specifications}
We estimate two gravity-type models at the city level to assess whether economic similarity exerts an influence on flow volumes after accounting for core geographic factors. Both are estimated in log-linear form using ordinary least squares.

\paragraph{(1). Baseline gravity model.}  
The baseline specification includes only origin and destination size effects (here measured by outflow and inflow volumes) and the standard distance decay term based on the geodesic distance \(d_{ij}\):
\begin{equation}
    F_{ij} = k \cdot \frac{P_i^\alpha P_j^\beta}{d_{ij}^\gamma}.
\end{equation}

\paragraph{(2). Economic–distance Gravity Model}  
We then extend the model by incorporating the income distance between two zones:
\begin{equation}
    F_{ij} = k \cdot \frac{P_i^\alpha P_j^\beta}{d_{ij}^\gamma} \cdot S(i, j),
\end{equation}

Here, $P_i$ and $P_j$ denote total outflow of origin tract $i$ and total inflow of destination tract $j$, respectively; $d_{ij}$ is the geodesic distance between tract $i$ and $j$. $k$ is a scaling constant, and $\alpha, \beta, \gamma$ are sensitivity coefficients for origin size, destination size and geographic distance. The economic-distance adjustment term is defined as
$S(i, j) = \exp\left(\theta \cdot |q_i - q_j|\right)$,
where $q_i$ and $q_j$ are income quantiles of tract $i$ and $j$, and $\theta$ is a free parameter to be estimated from the data, governing how strongly income distance affects mobility flows. Negative values of $\theta$ imply that larger income differences reduce mobility flows, corresponding to an income-based friction mechanism. All variables are log-transformed for estimation, and the model is fitted via ordinary least squares under the log-linear specification.

\subsection{Sliding-window $Q(m)$ measure and cluster analysis}

Building on the income-ordered OD matrix described above, we introduce a sliding-window measure $Q(m)$ to capture how inter-tract mobility is concentrated along the income hierarchy. For a given income midpoint $m \in [0.125, 0.875]$ (evaluated at intervals of 0.05), we define a square window of width $0.25$ along the diagonal of the income-quantile plane:
\[
[m-0.125,\, m+0.125] \times [m-0.125,\, m+0.125],
\]
so that the window includes flows connecting tracts with similar income levels. The measure $Q(m)$ is computed as the fraction of normalized inter-tract flows whose origin and destination income quantiles fall inside this square:
\[
Q(m) = \sum_{(i,j) \in \Omega(m)} \hat{F}_{ij},
\]
where $\hat{F}_{ij}$ denotes the normalized flow between tract $i$ and $j$, and $\Omega(m)$ is the set of origin-destination pairs included in the square centered at $m$. Under a random-matching null model, in which all origin-destination pairs are equally likely, the expected value of $Q(m)$ equals the area of the square window:
$Q_{\mathrm{null}}(m) = 0.25^2 = 0.0625$.
Values above 0.0625 indicate that flows are disproportionately concentrated within a narrow income band around $m$.

To summarize the shape of the $Q(m)$ profile for each city, we extracted four descriptive features:
(1) endpoint intensity,
$
\frac{Q(0.125)+Q(0.875)}{2}$,
which captures the average concentration at the low- and high-income extremes;
(2) asymmetry,
$
\frac{Q(0.875)-Q(0.125)}
{Q(0.875)+Q(0.125)},
$
which measures whether localized mobility is more concentrated among high- or low-income neighborhoods;
(3) U-shape depth,$\frac{Q(0.125)+Q(0.875)}{2}-Q(0.5)$,
which quantifies the contrast between the income extremes and middle-income neighborhoods; and
(4) the overall mean $\bar{Q}$,
which reflects the average intensity of income-localized mobility across the entire income hierarchy.

We applied K-means clustering to these four features to identify distinct city-level mobility typologies. The optimal number of clusters was selected as $k=4$ based on cluster interpretability and stability. To visualize the clustering structure, we additionally projected the four-dimensional feature space onto two principal components using principal component analysis (PCA).

\section*{Acknowledgements}

This work is supported by the National Natural Science Foundation of China (Grant Nos. 72288101, 42361144718, 72271019) and the Scientific Research Foundation of Beijing Jiaotong University (Grant No. 2025XKBH003).




\section*{Competing interests}  
The authors declare no competing interests.

\newpage 










\begin{appendices}

\pagestyle{fancy} 
\setcounter{page}{1}  

\begin{center}
\vspace*{0.3cm} 
\Large\bfseries Supplementary Information
\vspace{0.3cm}
\end{center}
\addcontentsline{toc}{section}{Supplementary Information} 



\renewcommand{\thefigure}{S\arabic{figure}} 

\renewcommand{\thetable}{S\arabic{table}}
\setcounter{figure}{0}
\setcounter{table}{0}

\renewcommand{\thesection}{S\arabic{section}}
\renewcommand{\appendixname}{}
\startcontents
\printcontents{}{1}{\section*{Supplementary Note}}
\vspace{1cm}


\section{Data Description}
This study primarily utilizes three categories of data: mobile phone travel flow data, income data at the U.S. Census Tract (CT) level, and city-level socioeconomic indicators. Together, these datasets form the foundation for investigating the relationship between urban mobility structures and socioeconomic structures.

This study focuses on the 109 largest metropolitan areas by population in the United States, covering the contiguous 48 states and Alaska (excluding Hawaii). One city in Alaska (Anchorage) is included in the analysis but not shown in Fig.~1a (main text), which only displays the contiguous states. We collected mobile phone signaling data from these cities, recorded in the form of Origin–Destination (OD) matrices~\cite{kang2020multiscales}. At the urban scale, each Census Tract is treated as a basic spatial unit, and the OD matrices capture travel flows from origin CTs to destination CTs. Each individual was first assigned a home census tract based on their most frequently visited nighttime (6 pm–7 am local time) tract over a six-week observation window. When an individual travels from the home tract A to multiple destination tracts during a day (e.g., A $\xrightarrow{}$ B $\xrightarrow{}$ C), the dataset records two separate flows: A $\xrightarrow{}$ B and A $\xrightarrow{}$ C, with any repeated visits to the same tract within a day counted only once. In other words, all flows are defined with respect to the individual’s home tract as the origin, regardless of intermediate stops. Finally, the accessible dataset provides daily origin-destination (OD) matrices containing tract GEOIDs, tract geographical coordinates,
date stamps, and visitor counts. In this study, daily OD matrices from January 2019 through February 2020 were aggregated into a single cumulative flow matrix, representing the total inter‑tract mobility over the full observation window. Data after February 2020 were excluded to avoid potential confounding effects of the COVID‑19 pandemic on mobility patterns.

Concurrently, economic income data for each CT within the cities were obtained from the U.S. Census Bureau. To measure the socioeconomic status (SES) of different spatial units, this study employs the median household income at the Census Tract level as a key proxy variable for socioeconomic status. At the city level, a multidimensional set of socioeconomic indicators was collected to characterize urban demographic structure, economic development level, and transportation patterns. These indicators mainly include the following categories:

(a) Population and Structure: Total population size, minority group proportion, racial diversity index, proportion of minors, and aging pressure index.

(b) Education and Employment: Proportion of population with higher education, proportion of high-skill occupations, and unemployment rate.

(c) Income and Poverty: Median household income, per capita income, income inequality (Gini coefficient), and poverty rate.

(d) Transportation Mode: Proportion of commuters driving alone, carpooling proportion, public transportation proportion, walking proportion, work-from-home proportion, and average commute time.

(e) Social Security: Health insurance coverage rate.

\section{Robustness of the Mobility Structure Index \(E(0.25)\)}
The choice of \(x=0.25\) is supported by two empirical patterns from the main text (Fig.~2b-c).The cumulative deviation \(E(x)-\tilde{E}(x)\) rises steeply for \(x<0.25\), then enters a plateau between \(x\approx0.25\) and \(x\approx0.35\). Within this plateau, the slope of the deviation curve is close to zero, meaning that the actual cumulative flow \(E(x)\) increases at nearly the same rate as the benchmark \(\tilde{E}(x)\). Hence, the excess concentration accumulated up to \(x=0.25\) is not further augmented by including flows with income distances between \(0.25\) and \(0.35\). This indicates that the strongest overconcentration of flows occurs precisely within the \(x<0.25\) range, making \(x=0.25\) the most efficient threshold to capture the core signal of flow concentration. This pattern is also reflected in the derivative \(E'(x)\), which shows a pronounced peak below\(x=0.25\) (Fig.~2c). Thus, \(E(0.25)\) captures the core of the flow concentration signal.

To assess whether the choice of the specific threshold \(x=0.25\) influences our conclusions, we examined the consistency of \(E(0.25)\) with alternative thresholds (\(E(0.30)\) and \(E(0.35)\)) as well as with the area under the full cumulative curve (AUC). As shown in Table~\ref{tab:up-down-robustness}, all pairwise correlations are above \(0.95\) and highly significant (\(p<0.001\)), indicating that \(E(0.25)\) captures essentially the same cross-city variation as both adjacent thresholds and the overall shape of the \(E(x)\) curve. These results confirm that our primary mobility structure index is robust to threshold selection and reliably reflects the overall flow concentration across U.S. cities.

\begin{table}[htbp]
\centering
\caption{Robustness of the $E(0.25)$ Segregation Index}
\setlength{\tabcolsep}{12pt}  
\begin{tabular}{lccc}
\toprule
& \multicolumn{2}{c}{Correlation with $E(0.25)$} \\
\cmidrule(lr){2-3}
Variable & Pearson $r$ & Spearman $\rho$ \\
\midrule
$E(0.30)$ & 0.963 & 0.978 \\
$E(0.35)$ & 0.955 & 0.969 \\
AUC       & 0.955 & 0.976 \\
\bottomrule
\end{tabular}
\par\smallskip

\begin{tablenotes}
    \item Note: Pearson and Spearman correlation coefficients between $E(0.25)$ and alternative thresholds $E(0.30)$, $E(0.35)$, and the area under the curve (AUC). All correlations are significant at $p<0.001$.
\end{tablenotes}

\label{tab:up-down-robustness}
\end{table}

\section{Robustness and Consistency of Directional Mobility Structure Indices}

\begin{figure}[h]
    \centering
    \includegraphics[width=1\linewidth]{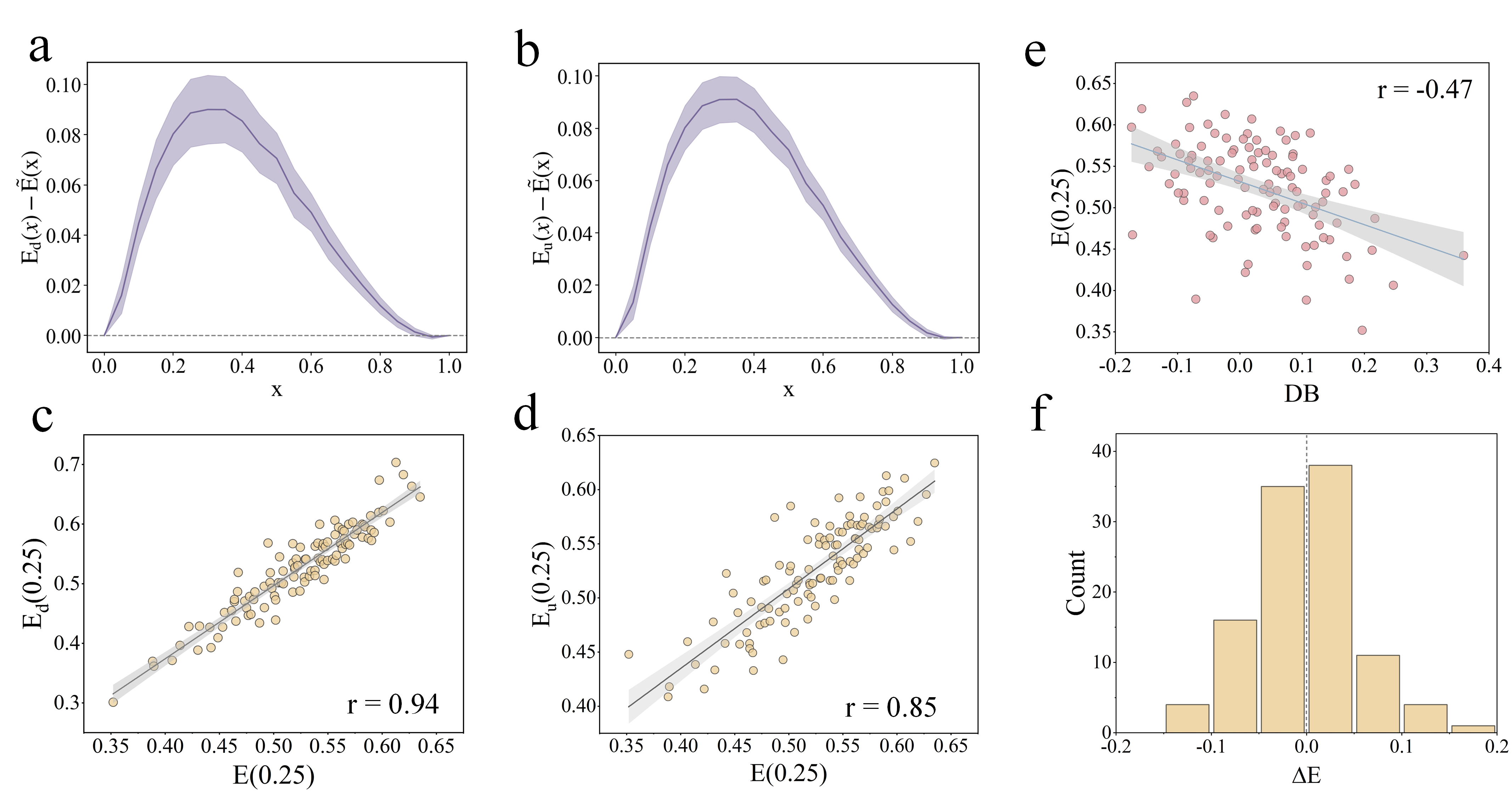}
    \caption[Directional Robustness and Consistency of Mobility Structure Measures.]{
\textbf{(a)} Mean deviation $E_{\mathrm{d}}(x)-\tilde{E}(x)$ for downward flows (higher$\rightarrow$lower). 
\textbf{(b)} Mean deviation $E_{\mathrm{u}}(x)-\tilde{E}(x)$ for upward flows (lower $\rightarrow$ higher). 
In both cases, the deviation peaks around $x \approx 0.25$–$0.35$, supporting the use of $x=0.25$ as a representative threshold. 
\textbf{(c)} Correlation between overall segregation $E(0.25)$ and downward segregation $E_{\mathrm{d}}(0.25)$ across cities. 
\textbf{(d)} Correlation between $E(0.25)$ and upward segregation $E_{\mathrm{u}}(0.25)$. 
Each point represents a city; lines show linear fits. 
\textbf{(e)} Scatter plot of directional bias (DB) versus $E(0.25)$. 
The correlation is moderate ($r=-0.47$, $p<0.001$), suggesting that cities with net downward dominance tend to have slightly higher overall segregation.
\textbf{(f)} Distribution of $\Delta E = E_{\mathrm{d}}(0.25)-E_{\mathrm{u}}(0.25)$ across cities. 
The approximately symmetric distribution indicates that directional differences in short-distance concentration are balanced across cities.
}
    \label{fig:up-down-index}
\end{figure}

To examine whether the choice of the threshold $x=0.25$ remains appropriate when mobility flows are decomposed by direction, we replicate the deviation analysis separately for downward (higher-income $\rightarrow$ lower-income) and upward (lower-income $\rightarrow$ higher-income) flows. Fig.~\ref{fig:up-down-index}a-b show the mean deviation $E(x)-\tilde{E}(x)$ for $E_{\mathrm{d}}(x)$ and $E_{\mathrm{u}}(x)$ across cities. In both cases, the deviation curves exhibit a clear peak in the range $x \approx 0.25$–$0.35$, consistent with the pattern observed for the aggregate function $E(x)$ in the main text. This confirms that the threshold $x=0.25$ captures the range where the overrepresentation of short economic-distance flows is most pronounced in both directional components. To further assess robustness, we compare $E_{\mathrm{u}}(0.25)$ and $E_{\mathrm{d}}(0.25)$ with alternative thresholds ($0.30$, $0.35$) and the area under the curve (AUC). As shown in Table~\ref{tab:directional_robustness}, all correlations remain high (Pearson $r>0.90$) and highly significant, indicating that the directional indices at $x=0.25$ capture consistent cross-city variation in mobility structure patterns. Correlations are slightly stronger for downward flows than for upward flows, reflecting the greater homogeneity of downward mobility structures across cities.

\begin{table}[htbp]
\centering
\caption{Robustness of Directional Mobility Structure Indices}
\setlength{\tabcolsep}{10pt}
\begin{tabular}{lcc}
\toprule
& Pearson $r$ & Spearman $\rho$ \\
\midrule
\multicolumn{3}{l}{\textit{Downward flows ($E_{\mathrm{d}}$)}} \\
\cmidrule(lr){1-3}
$E_{\mathrm{u}}(0.30)$ & 0.978 & 0.983 \\
$E_{\mathrm{u}}(0.35)$ & 0.978 & 0.977 \\
AUC                   & 0.970 & 0.974 \\
\midrule
\multicolumn{3}{l}{\textit{Upward flows ($E_{\mathrm{u}}$)}} \\
\cmidrule(lr){1-3}
$E_{\mathrm{d}}(0.30)$ & 0.939 & 0.954 \\
$E_{\mathrm{d}}(0.35)$ & 0.903 & 0.908 \\
AUC                   & 0.915 & 0.944 \\
\bottomrule
\end{tabular}
\par\smallskip

\begin{tablenotes}
    \item Note: Pearson and Spearman correlations between $E_{\mathrm{u}}(0.25)$ / $E_{\mathrm{d}}(0.25)$ and alternative thresholds ($0.30$, $0.35$) as well as the area under the curve (AUC). All correlations are significant at $p<0.001$.
\end{tablenotes}
\label{tab:directional_robustness}
\end{table}

We further assess the relationship between directional and overall flow concentration by correlating $E(0.25)$ with $E_{\mathrm{d}}(0.25)$ and $E_{\mathrm{u}}(0.25)$ across cities (Fig.~\ref{fig:up-down-index}c–d). Both directional indices are strongly positively correlated with the aggregate measure, with a higher correlation for downward flows ($r=0.94$) than for upward flows ($r=0.85$). This asymmetry indicates that cross-city variation in overall mobility structure is more closely associated with differences in downward mobility, whereas upward flow concentration remains relatively uniform across cities.

We further explore the relationship between directional dominance and overall mobility structure by examining the correlation between the directional bias index (DB, see Methods in the main text) and the aggregate measure $E(0.25)$ (Fig.~\ref{fig:up-down-index}e). The correlation is moderate ($r=-0.47$, $p<0.001$), indicating that cities with net downward dominance tend to have slightly higher overall flow concentration. 

To characterize directional asymmetry in short-distance concentration, we define $\Delta E = E_{\mathrm{d}}(0.25)-E_{\mathrm{u}}(0.25)$. As shown in Fig.~\ref{fig:up-down-index}f, $\Delta E$ is approximately symmetrically distributed around zero, indicating that neither direction systematically dominates in terms of short-distance flow concentration across cities. Importantly, $\Delta E$ captures the relative difference in short-distance clustering between the two directions, rather than the overall directional bias in flow volumes. As such, it provides a complementary perspective to the directional bias index (DB), which directly measures the net imbalance between upward and downward flows.

\clearpage

\section{Definitions of Residential Segregation Indicators}

In this study, we employ four widely used indicators to quantify the residential segregation underlying urban mobility patterns~\cite{Massey1988s}. Below we briefly describe each indicator and its computation.

\textbf{Evenness–centralization.} We compute the Dissimilarity Index ($D$) and the Entropy Index (Theil’s $H$) to capture macro‑scale unevenness in income distribution across tracts. Both indices are based on the full ten‑bin income distribution and follow standard formulations~\cite{reardon2011measuress}; $D$ ranges from 0 (perfect evenness) to 1 (complete segregation), while $H$ increases with greater mixing.

\textbf{Clustering–exposure.} To quantify spatial clustering of income levels, we compute Global Moran’s $I$ across a range of $k$-nearest neighbor definitions ($k=2$ to $22$)~\cite{getis2009spatials}. The $k$ value that maximizes the correlation with $E(0.25)$ is selected for the main analysis ($k=9$). We compute $I$ both with raw median income and with income quantiles; the quantile‑based version captures a neighborhood’s ordinal position in the city’s income hierarchy. Additionally, we employ Standardized Join Counts to test whether tracts of similar income category are more frequently adjacent than expected by chance~\cite{epperson2003covariancess}. Tracts are first classified into $L$ income categories based on city‑level quantiles. For the same $k$-nearest neighbor definition, we calculate the observed number of adjacent pairs ($O$) belonging to the same category, and compare it to the expected number under spatial randomness,
\begin{equation}
    R = M \cdot \sum_{c} (n_c / N)^2,
\end{equation}
where $R$ is the expected number of adjacent pairs belonging to the same income category under spatial randomness, $M$ is the total number of adjacent pairs, $n_c$ the number of tracts in category $c$ ($c\leq L$), and $N$ the total number of tracts. The standardized Join Count is then 
\begin{equation}
    Z = (O - R) / \sqrt{\text{Var}},
\end{equation}
with $\text{Var} = M \cdot \sum_{c} (n_c / N)^2 \cdot (1 - \sum_{c} (n_c / N)^2)$. Positive $Z$ indicates that same‑income tracts are more spatially clustered than expected. We evaluate the Join Count statistic over a range of income categories $L$ and neighborhood sizes $k$, and select the parameters that yield the strongest correlation with $E(0.25)$. This optimum occurs at $L=3$ and $k=10$ for the Join Count.

\section{Explaining Mobility Structure by Residential Stratification and Socioeconomic Characteristics}

We estimate a set of nested OLS regression models to examine the relative contributions of residential spatial clustering (Moran’s $I$) and income unevenness ($D$) to city-level mobility structure. All explanatory variables are standardized to facilitate comparison of coefficient magnitudes. The baseline specifications include univariate models for each predictor, followed by a bivariate model and a model with an interaction term. Standard errors are reported in parentheses.

While the two dimensions of residential stratification explain the majority of cross-city variation in urban mobility structure, a nontrivial share of variation remains unexplained. To examine whether this residual component exhibits systematic associations with broader urban and demographic characteristics, we use the residuals from column (4) in Table~1—i.e., the outcome after controlling for Moran’s $I$ and the Dissimilarity Index in the full model with interaction~\cite{lovell2008simples}. By construction, this residualized outcome is orthogonal to both Moran’s $I$ and the Dissimilarity Index, allowing us to isolate variation in mobility structure that cannot be attributed to residential spatial patterns alone. This residual variation further corresponds to uneven socioeconomic stratification emergent from mobility processes.

\begin{table}[htbp]
\centering
\caption{Residual Regression: Explaining Mobility Structure Beyond Residential Stratification}
\label{tab:stage2}
\begin{tabular}{lc}
\toprule
 & Residual mobility structure \\
\midrule
Constant                         & $0.000$ \\
                                 & (0.003) \\
Commute Public Transit Share     & $-0.0073$ \\
                                 & (0.005) \\
Average Commute Time             & $0.0069$ \\
                                 & (0.005) \\
Higher Education Share           & $0.0015$ \\
                                 & (0.005) \\
Income Inequality (Gini)         & $-0.0008$ \\
                                 & (0.004) \\
Population Share Under 18        & $-0.0114^{***}$ \\
                                 & (0.004) \\
Racial Diversity Index           & $0.0086^{***}$ \\
                                 & (0.003) \\
Total Population (2020)          & $0.0074$ \\
                                 & (0.009) \\
\midrule
$R^{2}$                          & 0.309 \\
Adjusted $R^{2}$                 & 0.261 \\
AIC                              & $-487.3$ \\
BIC                              & $-465.8$ \\
RMSE                             & 0.0240 \\
Observations                     & 109 \\
\bottomrule
\end{tabular}

\begin{tablenotes}
\footnotesize
\item Notes: Dependent variable is the residual from Table~1, Column~(4) (full model with interaction). All explanatory variables are standardized. Heteroskedasticity-robust (HC3) standard errors in parentheses. * $p<0.1$, ** $p<0.05$, *** $p<0.01$.
\end{tablenotes}
\end{table}

Table~\ref{tab:stage2} reports the results of this residual regression. All explanatory variables are standardized to facilitate comparison of effect magnitudes, and heteroskedasticity-robust (HC3) standard errors are applied. The model explains approximately 31\% of the residual variance in urban mobility structure ($R^{2}=0.309$), a meaningful improvement over previous specifications, and is statistically significant ($F$-test $p<0.001$), confirming that the residual variation exhibits systematic structural patterns.

Two covariates show statistically significant associations with residual mobility structure. First, racial diversity exhibits a positive and significant association ($\beta=0.0086$, $p=0.006$), indicating that in racially diverse cities, income-based flow concentration is stronger than predicted by residential spatial patterns alone. This suggests that race-related social boundaries may intersect with and amplify economic sorting in daily movement, ultimately reinforcing urban socioeconomic stratification. Second, the share of population under age~18 shows a negative and significant association ($\beta=-0.0114$, $p=0.003$), implying that family-oriented urban contexts tend to have weaker-than-predicted economic flow concentration, possibly due to more localized activity patterns driven by child-related travel. 

All other covariates--including public transit use, average commute time, educational attainment, income inequality (Gini coefficient), and city population—remain statistically insignificant. In particular, the lack of association with income inequality reinforces that its influence on urban mobility structure operates primarily through its effect on residential sorting rather than through an independent behavioral channel. Taken together, these results indicate that structural mobility patterns beyond residential spatial stratification exhibit systematic, but secondary, associations with cities’ demographic composition, reinforcing the primacy of residential spatial structure as the dominant driver of economic-distance-structured mobility.

\section{Directional Mobility Structure Regression Results}
This section presents supplementary regression results for upward ($E_\text{u}(0.25)$) and downward ($E_\text{d}(0.25)$) directional mobility structure, which validate the divergent mechanistic drivers of directional asymmetry in economic-distance-structured mobility (Table~\ref{tab:si_directional_reg}).

\textbf{Residential Stratification Drivers.} We first estimate the effects of residential spatial structure (Moran’s $I$ for spatial clustering and the Dissimilarity Index $D$ for income evenness) on directional mobility structure. Consistent with the core findings for overall mobility structure ($E(0.25)$), both spatial clustering and income dissimilarity exhibit significant positive associations with both $E_\text{u}(0.25)$ and $E_\text{d}(0.25)$. However, the magnitude of effects differs substantially across directions: spatial clustering (Moran’s $I$) exerts a stronger influence on downward mobility structure ($\beta=0.0439$, $p<0.001$) than on upward mobility structure ($\beta=0.0254$, $p<0.001$), and the income dissimilarity index $D$ also shows a larger coefficient for $E_\text{d}(0.25)$ ($\beta=0.0206$, $p<0.001$) relative to $E_\text{u}(0.25)$ ($\beta=0.0089$, $p<0.05$). A significant negative interaction term between Moran’s $I$ and $D$ is only observed for $E_\text{u}(0.25)$ ($\beta=-0.0070$, $p<0.05$), indicating a substitutive relationship between spatial clustering and income evenness in shaping upward mobility barriers—an effect absent for downward mobility. Model fit metrics confirm that residential stratification explains a larger share of variance in $E_\text{d}(0.25)$ (adjusted $R^2=0.6580$) than in $E_\text{u}(0.25)$ (adjusted $R^2=0.4979$), reinforcing that downward flow structure is more tightly linked to residential spatial patterns.

\textbf{Residual Analysis: Socioeconomic Correlates.} We further examine the socioeconomic correlates of directional mobility structure by focusing on the variation that is orthogonal to residential stratification. Specifically, we use the residuals from the baseline specification—after partialling out Moran’s $I$ and the Dissimilarity Index $D$—as the dependent variable.
The residual variation exhibits dimensionally distinct patterns for upward and downward mobility:
(i) For $E_\text{u}(0.25)$, residuals are weakly associated with urban spatial efficiency indicators, including public transit use ($\beta=-0.0092$, $p<0.1$) and average commute time ($\beta=0.0094$, $p<0.1$), with no significant links to socio-demographic characteristics (e.g., racial diversity, under-18 population share).
(ii) For $E_\text{d}(0.25)$, residuals are significantly linked to socio-demographic structure: racial diversity shows a positive association ($\beta=0.0103$, $p<0.1$), while the share of population under age 18 exhibits a negative association ($\beta=-0.0147$, $p<0.05$)--mirroring the residual correlates of overall mobility structure $E(0.25)$.

Other covariates (higher education ratio, income gini, city population) are not statistically significant for either direction, confirming that residual variation in directional mobility structure is weakly structured and context-specific. Collectively, these results demonstrate that directional asymmetry in economic-distance-structured mobility arises from divergent mechanistic pathways: upward mobility barriers are shaped by urban spatial efficiency, while downward mobility stickiness is modulated by socio-demographic context~\cite{Chetty2014s}.

\begin{table}[htbp]
\centering
\caption{OLS Regression Results for Directional Mobility Structure ($E_u(0.25)$ vs. $E_d(0.25)$)}
\begin{tabular}{lcc}
\toprule
\noalign{\vspace{-\aboverulesep}} 
\rowcolor{mycolor}  & (1) $E_\text{u}(0.25)$ & (2) $E_\text{d}(0.25)$ \\
\noalign{\vspace{-\belowrulesep}} 
\midrule
\multicolumn{3}{l}{\textbf{Stage 1: Residential Stratification Drivers (With Interaction)}} \\
\midrule
Constant & 0.5297$^{***}$ & 0.5285$^{***}$ \\
         & (0.004) & (0.005) \\
Moran's $I$ & 0.0254$^{***}$ & 0.0439$^{***}$ \\
          & (0.004) & (0.005) \\
$D$ (Income Dissimilarity) & 0.0089$^{**}$ & 0.0206$^{***}$ \\
                      & (0.004) & (0.005) \\
$I \times D$ & $-0.0070^{**}$ & -0.0045 \\
                                  & (0.003) & (0.004) \\
\midrule
$R^{2}$ & 0.5119 & 0.6675 \\
Adjusted $R^{2}$ & 0.4979 & 0.6580 \\
AIC & $-423.71$ & $-375.66$ \\
BIC & $-412.9$ & $-364.9$ \\
RMSE & 0.0334 & 0.0416 \\
Observations & 109 & 109 \\
\midrule
\multicolumn{3}{l}{\textbf{Stage 2: Residual Socioeconomic Correlates (HC3 SE)}} \\
\midrule
Constant & 1.045e-16 & 8.774e-17 \\
         & (0.003) & (0.004) \\
Commute Public Ratio & $-0.0092^{*}$ & -0.0057 \\
                     & (0.006) & (0.007) \\
Avg Commute Time & $0.0094^{*}$ & 0.0061 \\
                 & (0.005) & (0.007) \\
Higher Education Ratio & 0.0076 & -0.0023 \\
                       & (0.005) & (0.006) \\
Income Gini & -0.0037 & 0.0033 \\
            & (0.005) & (0.006) \\
Under18 Ratio & -0.0063 & $-0.0147^{**}$ \\
              & (0.005) & (0.006) \\
Racial Diversity Index & 0.0048 & $0.0103^{*}$ \\
                       & (0.005) & (0.005) \\
2020 Population & 0.0074 & 0.0069 \\
                & (0.007) & (0.012) \\
\midrule
$R^{2}$ & 0.2247 & 0.2268 \\
Adjusted $R^{2}$ & 0.171 & 0.173 \\
RMSE & 0.0294 & 0.0366 \\
Observations & 109 & 109 \\
\bottomrule
\end{tabular}
\begin{tablenotes}
\footnotesize
\item Note: All explanatory variables are standardized. Standard errors in parentheses (heteroscedasticity robust HC3 for Stage 2). * $p<0.1$, ** $p<0.05$, *** $p<0.01$.
\end{tablenotes}
\label{tab:si_directional_reg}
\end{table}

\clearpage
\section{Group-specific OLS Regression Results}

Table~\ref{tab:hetero_ols} reports the full group-specific OLS regression results corresponding to Fig.~5d–f in the main text. For each subgroup defined by city size, income inequality, and public transit usage, we estimate the model
\[
E(0.25) = \beta_0 + \beta_1 \cdot \text{Moran’s } I + \beta_2 \cdot D + \beta_3 \cdot (I \times D)
\]
using standardized predictors. The table presents coefficient estimates, statistical significance, model fit ($R^2$), and sample sizes.

Beyond the summary patterns discussed in the main text, the full results for the three core dimensions—city size, income inequality, and public transit usage—reveal several additional structural regularities. First, the decline in explanatory power across subgroups is systematic rather than incidental. Moving from low- to high-complexity urban contexts (small to large cities, low to high inequality, and low to high transit usage), the $R^2$ consistently decreases, indicating that residential spatial structure explains progressively less variation in mobility structure. Second, the relative importance of spatial clustering and income unevenness shifts across contexts. In small and medium cities as well as low-inequality settings, Moran’s $I$ remains the dominant predictor, with larger coefficients and higher statistical significance. In contrast, in high-inequality cities, the coefficient of $D$ surpasses that of Moran’s $I$, confirming a transition toward income-based stratification as the primary organizing mechanism of urban flow patterns. Third, the interaction term $(I \times D)$ exhibits context-dependent behavior. It is consistently negative in most groups, suggesting a substitutive relationship between spatial clustering and income unevenness. However, this interaction becomes statistically insignificant in large cities and high-transit environments, indicating a weakening of the coupling between spatial and economic residential stratification under more complex or better-connected urban conditions. Finally, the intercept terms increase systematically across high-group cities (e.g., large cities and high-transit cities), consistent with the higher baseline levels of concentrated flow patterns observed in the descriptive analysis. This pattern further supports the interpretation that intensified urban conditions are associated not only with structural shifts in drivers but also with elevated baseline flow concentration, which subsequently generates stronger socioeconomic stratification.

Together, these results provide detailed statistical support for the structural transition described in the main text: from spatially constrained mobility systems toward more complex regimes in which income unevenness gains prominence and overall predictability declines. These patterns are robust to alternative specifications without the interaction term (not shown), yielding qualitatively similar shifts in coefficient magnitudes and model fit.

We further examine three additional contextual factors: population density, racial diversity, and aging pressure. For these attributes, the pattern reverses: high‑group cities exhibit higher $R^2$ than low‑group cities (Table~\ref{tab:hetero_ols}), indicating that these factors tighten the coupling between residential structure and mobility structure. Specifically:

- Population density: High‑density cities show a significant negative interaction ($\beta=-0.017$, $p<0.05$) and a high $R^2=0.801$, suggesting that spatial compactness reinforces the substitutive relationship between spatial clustering and income unevenness, making mobility structure more dependent on neighborhood residential conditions.

- Racial diversity: High‑diversity cities also display elevated $R^2=0.807$, with the interaction term no longer significant. This may reflect that racial and income dimensions intersect, yet residential spatial structure remains a strong predictor of city-level mobility structure in racially diverse contexts.

- Aging pressure: High‑aging cities show a significant negative interaction ($\beta=-0.011$, $p<0.05$) and a substantial increase in $R^2$ (from $0.677$ to $0.815$). This likely reflects that older populations have more localized activity spaces, making daily mobility structure more tightly bound to the income composition of nearby neighborhoods.

These contrasting patterns underscore that urban characteristics do not uniformly modulate the role of residential spatial structure. While city size, inequality, and transit usage decouple mobility structure from residential patterns, factors like density, diversity, and aging pressure strengthen that link—likely by compressing activity spaces or reinforcing local socioeconomic dependence.

\begin{table}[!htbp]
\centering
\caption{Group-specific OLS Results for Mobility Structure $E(0.25)$}
\label{tab:hetero_ols}
\begin{tabular}{lccccccc}
\toprule
Group & Moran's $I$ & $D$ & Interaction & Constant & $R^2$ & $N$ \\
\midrule
Full sample      & $0.0347^{***}$ & $0.0149^{***}$ & $-0.0067^{*}$  & $0.5273$ & $0.7264$ & $109$ \\
Large cities     & $0.0202^{**}$  & $0.0139^{*}$   & $-0.0067$      & $0.5466$ & $0.4153$ & $36$  \\
Small/med cities & $0.0361^{***}$ & $0.0145^{***}$ & $-0.0044$      & $0.5191$ & $0.7786$ & $73$  \\
High Gini inequality  & $0.0244^{*}$   & $0.0274^{**}$  & $-0.0131^{*}$  & $0.5264$ & $0.5791$ & $36$  \\
Low Gini inequality   & $0.0370^{***}$ & $0.0113^{**}$  & $-0.0057^{.}$  & $0.5263$ & $0.8029$ & $73$  \\
High pub. transit     & $0.0348^{***}$ & $0.0175^{**}$  & $-0.0046$      & $0.5404$ & $0.6770$ & $36$  \\
Low pub. transit      & $0.0344^{***}$ & $0.0121^{*}$   & $-0.0081^{*}$  & $0.5213$ & $0.7728$ & $73$  \\
High density          & $0.0333^{***}$ & $0.0183^{***}$ & $-0.0171^{**}$ & $0.5267$ & $0.8007$ & $36$ \\
Low density           & $0.0334^{***}$ & $0.0147^{***}$ & $-0.0042$      & $0.5290$ & $0.7129$ & $73$ \\
High racial diversity & $0.0341^{***}$ & $0.0174^{***}$ & $-0.0075$      & $0.5280$ & $0.8067$ & $36$ \\
Low racial diversity  & $0.0345^{***}$ & $0.0144^{***}$ & $-0.0062^{*}$  & $0.5245$ & $0.6939$ & $73$ \\
High aging pressure   & $0.0312^{***}$ & $0.0152^{**}$  & $-0.0108^{**}$ & $0.5294$ & $0.8148$ & $36$ \\
Low aging pressure    & $0.0340^{***}$ & $0.0138^{***}$ & $-0.0043$      & $0.5251$ & $0.6769$ & $73$ \\
\bottomrule
\multicolumn{7}{l}{\footnotesize \textit{Note:} Significance levels: $^{.}p<0.1$, $^{*}p<0.05$, $^{**}p<0.01$, $^{***}p<0.001$.} \\
\end{tabular}
\end{table}

\section{Economic-distance Gravity Model Performance by Flow Direction}

To validate the economic-distance gravity model (EGM) relative to the baseline gravity model (BGM), we separately modeled upward (from lower- to higher-income neighborhoods) and downward (from higher- to lower-income neighborhoods) flows for each city. For each directional flow, we fit both models and compared their performance using two metrics: the coefficient of determination ($R^2$) and the reconstruction error (RMSE) of the corresponding cumulative flow curves ($E(x)$).

\begin{figure}[!h]
    \centering
    \includegraphics[width=0.8\linewidth]{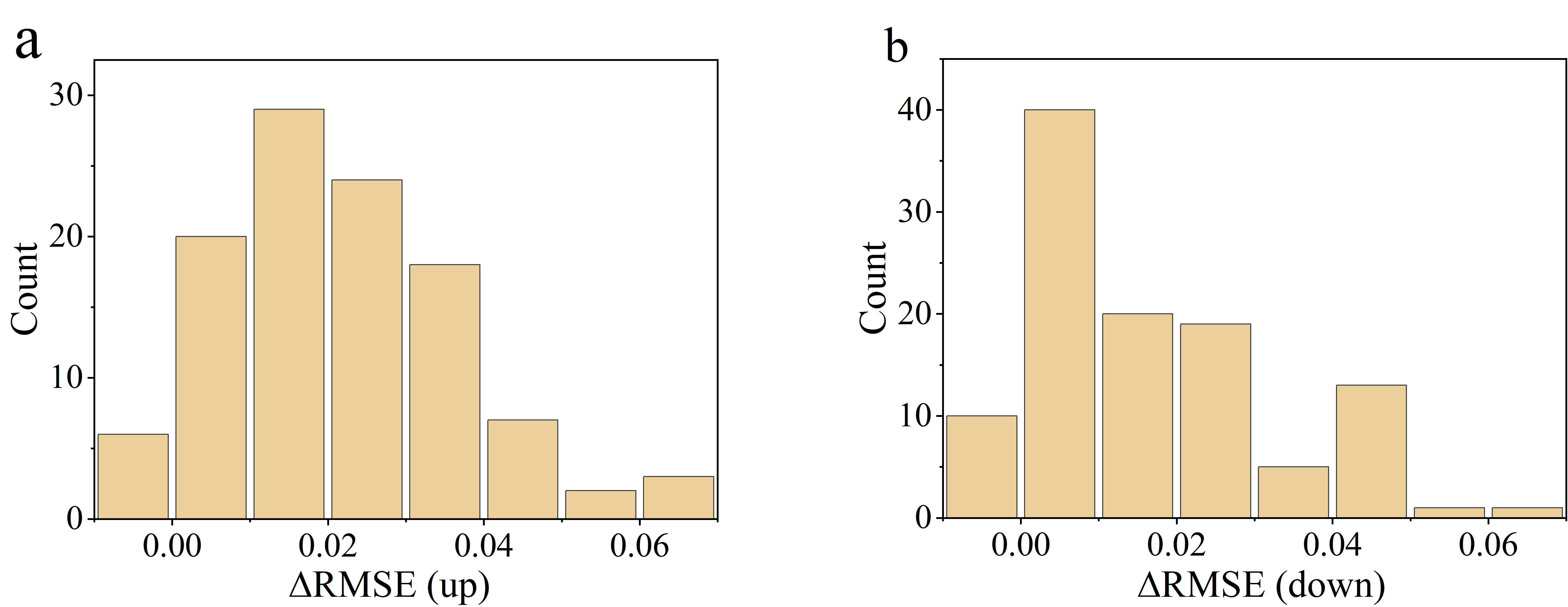}
    \caption[Economic-distance Gravity Model Performance Across Flow Directions]{
\textbf{(a-b)} Distribution of $\Delta RMSE = RMSE(BGM) - RMSE(EGM)$ for upward (a) and downward (b) flow cumulative curves curves $E(x)$. Positive values indicate that the EGM better reproduces the observed curves. Most cities show positive improvements (right side of zero), with only a few exceptions, confirming the robustness of the EGM across both flow directions.}
    \label{fig:deltaRMSE}
\end{figure}

For both upward and downward flows, the improvement in reconstruction error is quantified as $\Delta RMSE = RMSE(BGM) - RMSE(EGM)$, where positive values indicate that the EGM better reproduces the observed $E(x)$ curves. As shown in Fig.~\ref{fig:deltaRMSE}a–b, the majority of cities exhibit positive $\Delta RMSE$ for both directions, with only a small number of cities (6–10) showing slightly negative values. Together, these results demonstrate that the EGM provides stable and generalizable improvements in both model fit and structural accuracy across upward and downward flows, supporting economic distance as a fundamental driver of mobility patterns.

\clearpage

\section{Sliding-window mobility measure $Q(m)$ and city typologies}

To illustrate the concentration of short-distance mobility along the income hierarchy, we applied a sliding-window measure $Q(m)$, which quantifies the fraction of inter-tract flows occurring within a square window of width $0.25$ along the diagonal of the income-quantile plane. This window approximately corresponds to the range of income distances identified in the main text ($x \leq 0.25$) as containing the majority of concentrated flows. Values of $Q(m)$ above the uniform expectation indicate that mobility is disproportionately localized among tracts of similar income.

\begin{figure}[h!]
    \centering
    \includegraphics[width=0.9\textwidth]{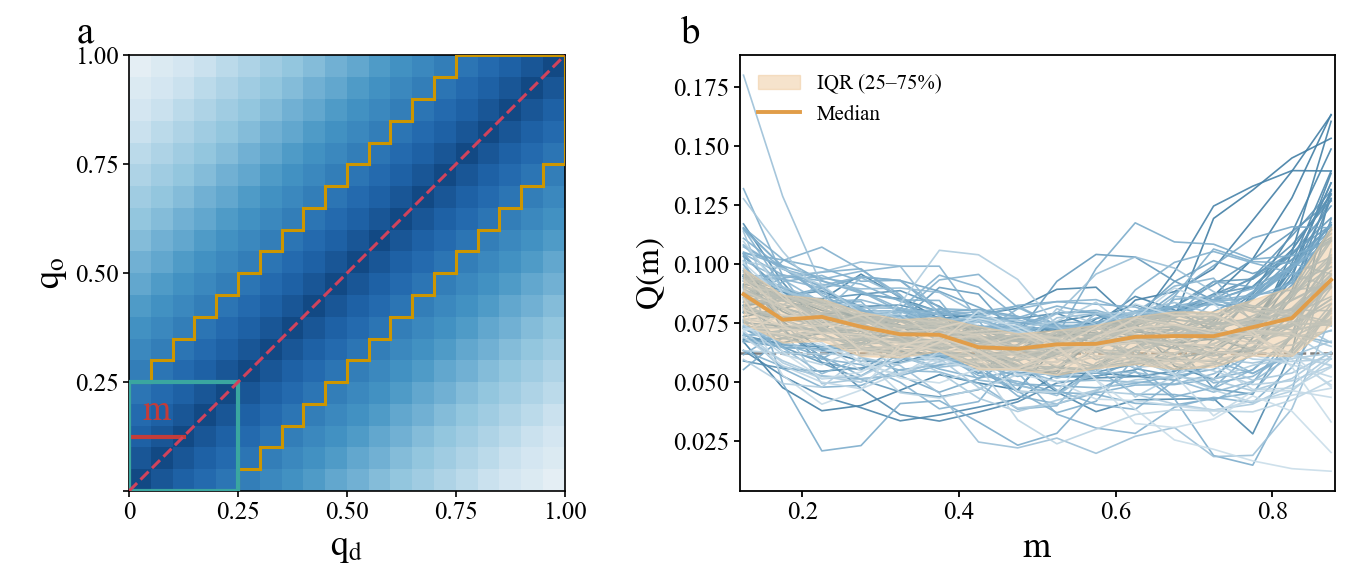}
    \caption[Illustration of $Q(m)$ measure and overall city patterns]{Illustration of $Q(m)$ measure and overall city patterns. 
    (\textbf{a}) Example schematic of a single city's normalized inter-tract flow matrix. The green square along the diagonal illustrates the sliding window (width 0.25) for a given $m$, from which $Q(m)$ is computed. 
    (\textbf{b}) Q curves for all cities, with yellow line showing the cross-city mean and shaded region representing the interquartile range (IQR), highlighting the characteristic U-shaped profile of internal mobility concentration.}
    \label{fig:Q_method_all}
\end{figure}

Figure~\ref{fig:Q_method_all}a shows an example of a single city's flow matrix with a green square illustrating the sliding window at one midpoint $m$. Figure~\ref{fig:Q_method_all}b presents the $Q(m)$ curves for all cities, along with the mean and interquartile range (IQR), revealing an overall U-shaped pattern of internal mobility concentration: both low- and high-income neighborhoods exhibit higher short-distance mobility compared with middle-income neighborhoods.

Cities were clustered based on four summary features derived from each $Q(m)$ curve: endpoint intensity (average $Q$ at low- and high-income extremes), asymmetry (relative concentration of high- vs. low-income neighborhoods), U-depth (difference between the endpoints and the middle), and overall mean $Q$. K-means clustering identified four distinct city types. Table~\ref{tab:cluster_features_by_type} reports the mean values of the four clustering features for each city type. 
To validate the K-means clustering results and visualize the separation of mobility regimes, we performed principal component analysis (PCA) on the four standardized features (endpoint intensity, asymmetry, U-depth and overall mean). All features were standardized using z-score transformation to eliminate scale effects. We retained the first two principal components for visualization, which collectively explain $85.79\%$ of the total variance. Convex hulls were plotted to highlight the spatial boundaries of each cluster in the reduced feature space. Notably, clustering was performed on the original four features, and PCA was used only for post-hoc visualization and validation, ensuring the typologies were defined by the intrinsic $Q(m)$ curve characteristics.

\begin{figure}[h!]
    \centering
    \includegraphics[width=1\textwidth]{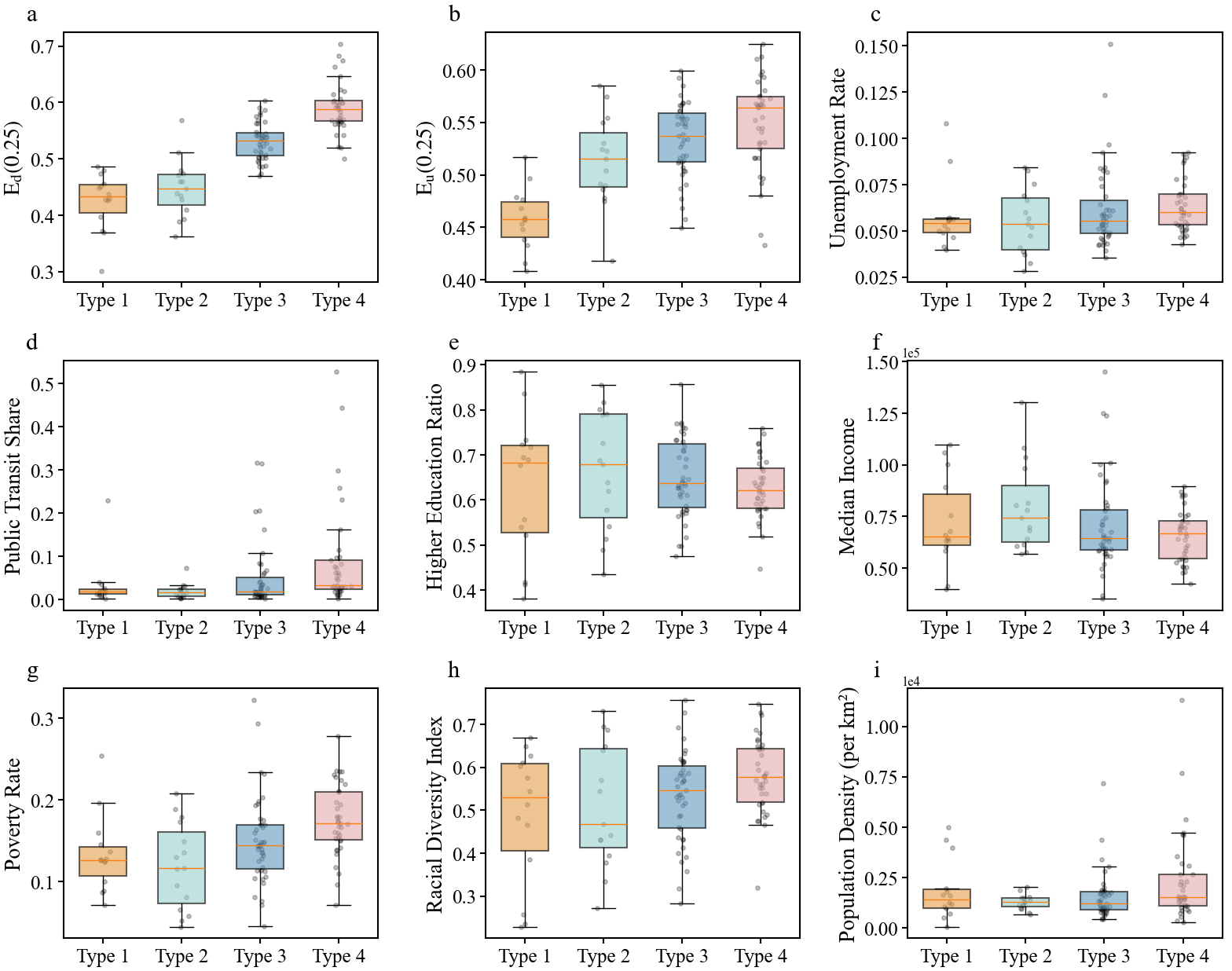}
    \caption[Distributions of selected city characteristics across the four mobility typologies]{Distributions of selected city characteristics across the four mobility typologies.
    (\textbf{a}) Downward mobility concentration ($B_\text{down\_025}$), (\textbf{b}) Upward mobility concentration ($B_\text{up\_025}$), (\textbf{c}) Unemployment rate, (\textbf{d}) Public transit usage, (\textbf{e}) Proportion of higher-education adults, (\textbf{f}) Median income, (\textbf{g}) Poverty rate, (\textbf{h}) Racial diversity, (\textbf{i}) Population density (km$^2$). 
    Boxplots indicate the median, interquartile range, and 1.5× IQR whiskers for each type, showing systematic differences between mobility typologies.}
    \label{fig:Q_boxplots}
\end{figure}

Figure~\ref{fig:Q_boxplots} displays the distributions of selected urban characteristics across the four mobility typologies. Panels include downward and upward mobility concentrations ($B_\mathrm{down\_025}$, $B_\mathrm{up\_025}$), unemployment rate, public transit usage, higher-education ratio, median income, poverty rate, racial diversity index, and population density. Systematic differences across typologies highlight that mobility concentration patterns align with broader socio-economic and demographic characteristics.

\begin{table}[htbp]
\centering
\caption{Mean values of clustering features by city type}
\label{tab:cluster_features_by_type}
\begin{tabular}{cccccc}
\toprule
Type & Number of cities & Endpoint mean & Asymmetry & U-depth & $\bar{Q}$ \\
\midrule
1 & 14 & 0.062 & -0.255 & 0.000 & 0.061 \\
2 & 15 & 0.095 & -0.150 & 0.049 & 0.060 \\
3 & 43 & 0.087 & 0.016 & 0.019 & 0.075 \\
4 & 37 & 0.108 & 0.183 & 0.046 & 0.079 \\
\bottomrule
\end{tabular}
\end{table}

\end{appendices}
\end{document}